\documentclass[aps,prb,twocolumn,groupedaddress,showpacs,tighten,eqsecnum]{revtex4}

\usepackage[dvips]{color} 
\usepackage{graphicx}
\usepackage{bm}      
\usepackage{latexsym}
\usepackage{amsmath,amssymb}
\usepackage{color}
\usepackage{stackrel}
\usepackage{accents}
\usepackage{mathrsfs}

\DeclareMathOperator{\sgn}{sgn}
\newcommand{\ord}[1]{\bm{\mathit{O}}\left(#1\right)}

\newcommand{\intl}[1]{\int\limits_{#1}}
\newcommand{\vex}[1]{\bm{\mathrm{#1}}}

\newcommand{\ts}[1]{\textstyle{#1}}

\newcommand{\bsub}{\begin{subequations}}
\newcommand{\esub}{\end{subequations}}

\newcommand{\T}{\mathsf{T}}

\newcommand{\tr}{\mathsf{Tr}}
\newcommand{\Nabla}{\bm{\nabla}}

\newcommand{\kap}{\kappa}

\newcommand{\sigh}{\hat{\sigma}}
\newcommand{\sigb}{\hat{\bm\sigma}}

\newcommand{\kah}{\hat{\kappa}}

\newcommand{\muh}{\hat{s}}

\newcommand{\lf}{\mathcal{L}}
\newcommand{\rt}{\mathcal{R}}
\newcommand{\msf}[1]{\mathsf{#1}}
\newcommand{\LF}{\mathsf{L}}
\newcommand{\RT}{\mathsf{R}}

\newcommand{\tk}{\hat{\mathfrak{t}}_{\kappa}}

\newcommand{\q}{\hat{Q}}

\newcommand{\pup}[1]{{\scriptscriptstyle{({#1})}}}
\newcommand{\puprm}[1]{\scriptscriptstyle{(\mathrm{{#1}})}}
\newcommand{\pupsq}[1]{{\scriptscriptstyle{[{#1}]}}}

\newcommand{\hci}{H_{\mathrm{CI}}^{\pup{0}}}
\newcommand{\hcio}{H_{\mathrm{CI}}^{\pup{0,1}}}
\newcommand{\hcit}{H_{\mathrm{CI}}^{\pup{0,2}}}
\newcommand{\hai}{H_{\mathrm{AIII}}^{\pup{0}}}
\newcommand{\haio}{H_{\mathrm{AIII}}^{\pup{0,1}}}
\newcommand{\hait}{H_{\mathrm{AIII}}^{\pup{0,2}}}
\newcommand{\hdi}{H_{\mathrm{DIII}}^{\pup{0}}}
\newcommand{\hdio}{H_{\mathrm{DIII}}^{\pup{0,1}}}
\newcommand{\hdit}{H_{\mathrm{DIII}}^{\pup{0,2}}}
\newcommand{\hcii}{H_{\mathrm{CI}}^{\pup{I}}}
\newcommand{\haii}{H_{\mathrm{AIII}}^{\pup{I}}}
\newcommand{\hdii}{H_{\mathrm{DIII}}^{\pup{I}}}

\newcommand{\lamN}{\lambda_{{\scriptscriptstyle{N\!A}}}}
\newcommand{\e}{\varepsilon}
\newcommand{\WZNW}{S_{{\scriptscriptstyle{\mathrm{WZNW}}}}}
\newcommand{\hf}{\eta}

\begin{document}

\title{Topological protection, disorder, and interactions:\\ Survival at the surface of 3D topological superconductors}
\author{Matthew S.\ Foster}
\email{matthew.foster@rice.edu} 
\author{Hong-Yi Xie}
\author{Yang-Zhi Chou}
\affiliation{Department of Physics and Astronomy, 
	     Rice University, 
             Houston, 
             Texas 77005,
	     USA}

\date{\today}

\begin{abstract}
We consider the interplay of disorder and interactions upon the gapless surface states of 
3D topological superconductors. The combination of 
topology
and superconducting order
inverts the action of time-reversal symmetry, so that extrinsic time-reversal invariant surface perturbations
appear only as ``pseudomagnetic'' fields (abelian and non-abelian vector potentials, which couple
to spin and valley currents).
The main effect of disorder is to induce multifractal scaling in surface state wavefunctions. 
These critically delocalized, yet strongly inhomogeneous states renormalize interaction
matrix elements relative to the clean system. We compute the enhancement or suppression of interaction 
scaling dimensions due to the disorder exactly, using conformal field theory. 
We determine the conditions under which interactions remain irrelevant in the presence of disorder 
for symmetry classes AIII and DIII. In the limit of large topological winding numbers 
(many surface valleys), we show that the effective field theory takes the form of a Finkel'stein 
non-linear sigma model, augmented by the Wess-Zumino-Novikov-Witten term. The sigma model incorporates 
interaction effects to all orders, and provides a framework for a controlled perturbative expansion; 
the 
inverse 
spin or 
thermal
conductance is the small parameter. For class DIII we 
show that interactions are always irrelevant, while in class AIII there is a finite window of 
stability, controlled by the disorder. Outside of this window we identify new interaction-stabilized 
fixed points. 
\end{abstract}


\maketitle

\tableofcontents


\section{Introduction \label{Sec: Intro}}

\begin{figure*}
   \includegraphics[width=0.88\textwidth]{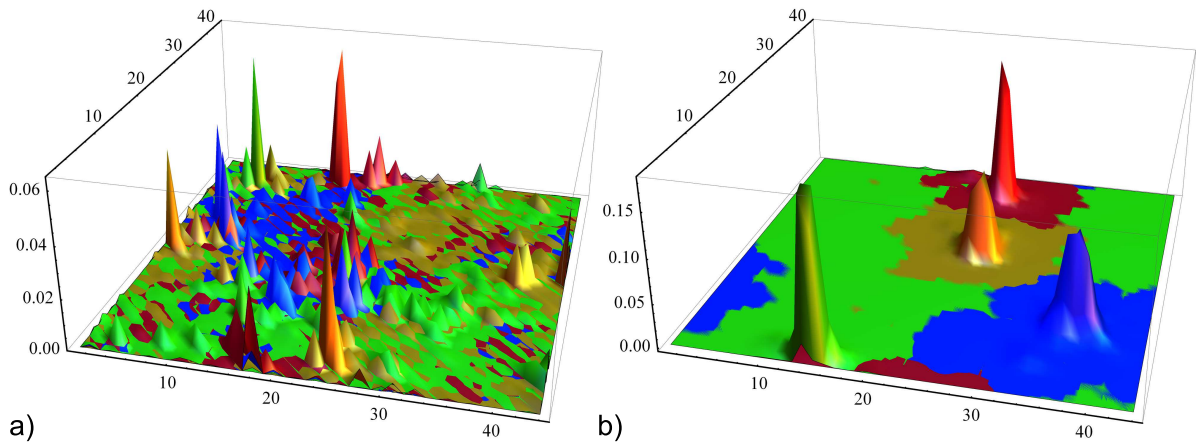}
   \caption{
	{\bf Multifractality, Chalker scaling, and interaction amplification.}
	Critically delocalized a) and Anderson localized b) states. 
	In each case we plot the probability density $|\psi_{n}(\vex{r})|^2$ for
	four different wavefunctions corresponding to four successive 
	energy eigenvalues $\e_n < \e_{n+1} < \e_{n+2} < \e_{n+3}$ 
	(blue, red, orange, green). Each critically delocalized
	state is multifractal,\cite{Paladin87,MFCRev,Loc}
	exhibiting an intricate inhomogeneous spatial
	structure. The structures of \emph{different} wavefunctions
	with nearby energy eigenvalues are strongly correlated, a phenomenon
	called Chalker scaling.\cite{Chalker88,Chalker90,Cuevas07} 
	This appears in a) as the interweaving of
	different colored peaks to form calico ``mountain ranges.'' 
	The combination of multifractality and Chalker scaling can 
	enhance short-ranged interparticle interactions.\cite{Feigelman07,Feigelman10,Foster12-B} 
	By contrast, for the Anderson localized states in b), 	
	there is essentially zero overlap between $|\psi_{n}(\vex{r})|^2$
	and $|\psi_{m}(\vex{r})|^2$ for states with 
	$0 < |\e_{n} - \e_{m}| \ll \delta_{l}$, so that 
	short-ranged interactions have negligible effect (at zero temperature).\cite{Basko06}
	Here $\delta_{l}$ denotes the level spacing in a characteristic
	localization volume. 
	These wavefunctions were computed for a 2D class BDI lattice 
	random hopping model,\cite{Loc} which is a non-topological analog
	for the CI surface states of a 3D topological superconductor.\cite{footnote--BDI}
	}
   \label{Fig--ChalkerWF}
\end{figure*}

The discoveries of graphene, 2D and 3D topological insulators have cemented ``Dirac materials'' as a
new physics frontier. Much recent interest has pivoted to employing these as platforms for
topological superconductivity and Majorana fermion zero modes, usually through proximity 
coupling to an ordinary s-wave superconductor.\cite{Alicea12} On the other hand, the discovery of \emph{bulk}
3D topological superconductors\cite{SRFL08,Kitaev09,SRL09,SRFL10,TISC,Roy08,Qi09,Fu10,Schnyder10,Yada11,Yamakage12,Xiang12,Hsieh12,Foster12-B} 
would give access to a fundamentally different type
of Dirac material, with features not easily attainable elsewhere.   

Bulk topological superconductors (TSCs) should solve the energy
mismatch problem that often plagues other Dirac materials. 
In a topological insulator, the chemical potential may reside
in the gap, in the valence or conduction bands, and the same is true
of the surface state Dirac point. In general there is no fundamental 
mechanism to pin the chemical potential in the gap and/or align it to the
Dirac point.\cite{TISC} 
TSCs are predicted to host Majorana bands of topologically-protected 
quasiparticles at the material surface.\cite{Volovik,SRFL08,TISC} 
These surface bands penetrate the bulk superconducting gap, and the 
chemical potential appears precisely at the Dirac point of the surface band 
(inside the gap), due to particle-hole symmetry.

Bulk TSCs can be realized in three of the
five topological classes of the Altland-Zirnbauer scheme
in three dimensions; these are denoted
DIII, AIII, and CI.\cite{SRFL08,Kitaev09,SRFL10} In order for these to be topological, 
time-reversal symmetry must be preserved in the bulk. The three classes 
differ by the degree of spin rotational symmetry. Class DIII has no spin symmetry, 
as occurs if strong spin-orbit coupling is present.\cite{Fu10,Yada11,Yamakage12,Xiang12,Hsieh12}
Class AIII has spin U(1) symmetry, as might arise in a 
time-reversal invariant spin-triplet p-wave superconductor.
Class CI has full spin SU(2) symmetry and spin singlet pairing.\cite{SRL09,Schnyder10,Foster12-B}
 
A remarkable feature of topological superconductors is that the effects
of time-reversal symmetry are ``inverted'': external perturbations of the
surface that respect time-reversal invariance
appear as pseudomagnetic fields (abelian or non-abelian vector potentials).\cite{SRFL08}
Because defects and disorder are inevitable at material interfaces,
a generic TSC surface will realize a kind of quenched (2+1)-D quantum 
chromodynamics, wherein massless Dirac quasiparticles navigate a 
landscape of frozen gauge fluctuations.\cite{BL02,SRFL08,Foster12-B}
(There is one exception, which is a single Dirac valley in class DIII.\cite{Volovik}
In this case there are no conserved currents 
except energy, and disorder 
can enter only through a modulation of the Fermi velocity.\cite{Nakai14})
The physics underlying this unusual version of time-reversal symmetry
is the unique conflagration of superconducting order and topology:
surface vector potentials couple to time-reversal even spin and valley
currents. The corresponding densities are odd. 
Topological protection forbids the appearance of a Dirac mass.

Equally remarkable is the response of the surface quasiparticles 
to non-magnetic disorder. Topological protection implies that at least
one surface wavefunction must remain delocalized.
As first demonstrated 20 years ago,\cite{Ludwig94,Nersesyan94} the problem of 2D Dirac fermions coupled
to quenched random (abelian or non-abelian) vector potentials is solvable
via conformal field theory. Wavefunctions near zero energy are critically delocalized,
and their universal statistical properties 
such as the multifractal\cite{Paladin87} spectrum
of local density of states fluctuations\cite{MFCRev,Loc}
can be computed 
exactly.\cite{Ludwig94,Chamon96,Castillo97,Carpentier01,Motrunich02,Horovitz02,Mudry03,Nersesyan94,Tsvelik95,Bernard95,Mudry96,Caux96,Ludwig00,Bhaseen01,LeClair08,Foster12-B} 

In this paper, we compute universal properties of and construct global 
phase diagrams for topological superconductor surface states.
We consider the simultaneous effects of both disorder and interactions.
We determine properties of the critically delocalized surface states in 
the absence of interactions, using conformal field theory.\cite{Ludwig94,Nersesyan94,Mudry96,Caux96,Chamon96,Ludwig00,Bhaseen01,LeClair08,Foster12-B}
Our results for the class DIII multifractal spectra are new.
In our previous work\cite{Foster12-B} we analyzed the effects of interactions
on class CI TSCs, which possess spin SU(2) invariance.
Here
we do the same for classes AIII and DIII, which respectively possess $U(1)$ and no spin symmetry. 
We enumerate 
four fermion interactions consistent with bulk symmetries, 
and compute the enhancement or suppression of interaction matrix elements due to the disorder.
The combination of wavefunction multifractality\cite{MFCRev,Loc} and dynamical 
Chalker scaling\cite{Chalker88,Chalker90,Cuevas07} can amplify interactions\cite{Feigelman07,Feigelman10}
and even sabotage topological protection, as we found in class CI.\cite{Foster12-B} 
See Figure~\ref{Fig--ChalkerWF}.

We show that in the limit of large topological winding numbers (many surface valleys),
a weakly-coupled sigma model allows a complimentary perturbative approach.
We graft interparticle interactions into this framework, and thereby
obtain Wess-Zumino-Novikov-Witten\cite{CFT} Finkel'stein non-linear sigma models\cite{Finkelstein,BK} (WZNW-FNLsMs) for all
three classes. Using the exact conformal field theory results and 
one-loop WZNW-FNLsM 
renormalization group (RG)
equations,\cite{Foster06,DellAnna06,Foster08,Foster12-B,WZNW-P3}
we establish the restrictive conditions under which surface states can be 
robust to both disorder and interactions.
For class DIII we show that interactions are always irrelevant, while in 
class AIII there is a finite window of stability, controlled by the disorder. Outside of this window 
we identify new interaction-stabilized fixed points. 

We stress that here we focus on \emph{weak} disorder and \emph{weak} interactions.
In particular, we consider only the effects of disorder at the surface, and we assume that 
the energy bandwidth of the microscopic disorder distribution is narrower than the bulk gap.
Sufficiently strong surface disorder can induce a topologically-trivial Anderson insulator in the outer
layers of the crystal, submerging the non-trivial surface bands below.\cite{Schubert12}
Strong interactions at the surface can induce spontaneous time-reversal symmetry breaking, or
realize gapped phases with topological order.\cite{Fidkowski13,Wang13,Wang14} 
For modest disorder or interactions, it has been argued that the surface states of topological
insulators and superconductors are protected.\cite{TISC} We have shown that this
is false in class CI, wherein we argued that the combination of 
arbitrarily weak disorder 
\emph{and} 
interactions
always induces spontaneous time-reversal symmetry breaking at surface.\cite{Foster12-B}

Compared to analyzing the interplay of disorder and interactions in other Dirac materials,\cite{Ostrovsky10,Nandkishore13}
the task of understanding topological superconductor surface states is much easier, due 
to the availability of non-perturbative techniques.
The key tools employed in this work are conformal embeddings.\cite{CFT}
A conformal embedding defines systematic ``fractionalization rules'' for
breaking up a level one affine Lie algebra (such as free fermions) into
generalized ``color'' and ``flavor'' sectors.
This is the non-abelian version of spin-charge separation that 
applies e.g.\ to (1+1)-D quantum chromodynamics.\cite{QCD}
For Dirac fermions coupled to quenched gauge fluctuations,
the relevant embeddings are the three infinite two-parameter families\cite{Affine}
\begin{align}\label{Embed}
\begin{array}{llll}
	\textrm{Class } & \textrm{Embedding} & |\nu|  & \\
\hline
	\textrm{CI} & \textrm{SO}(4 n k)_1 \supset \textrm{Sp}(2 n)_k \oplus \textrm{Sp}(2 k)_n \; & 2 k &(k \geq 1) \\
	\textrm{AIII} & \textrm{U}(n k)_1 \supset \textrm{U}(n)_k \oplus \textrm{SU}(k)_n \; & k &(k \geq 2) \\
	\textrm{DIII} & \textrm{SO}(n k)_1 \supset \textrm{SO}(n)_k \oplus \textrm{SO}(k)_n \; & k  &(k \geq 3)\\
\hline
\end{array}
\end{align}
Each family applies to a particular 3D TSC symmetry class.
In this table, $|\nu|$ is the modulus of the bulk winding number,
equal to the number of valleys (``colors''). 
For each embedding, this is specified by the parameter $k$,
while the parameter $n$ plays the role of a replica (``flavor'') index. 
The replica index is used to perform disorder-averaging.\cite{Nersesyan94,Tsvelik95,Caux96,AltlandSimons} 
Identical results 
have been obtained for multifractal spectra in class AIII\cite{Mudry96} and 
dynamic critical exponents in all three classes\cite{LeClair08}
using the supersymmetric\cite{SUSY} method; the corresponding embeddings appear in Ref.~\onlinecite{LeClair08}.

Three special cases do not fall into the scheme of Eq.~(\ref{Embed}).
The minimal realization of class AIII has one surface valley ($|\nu| = 1$).
In this case, time-reversal invariant perturbations couple only to the
$U(1)$ spin current, in the form of an abelian vector potential.\cite{Ludwig94,SRFL08,YZC-P1}
The minimal $|\nu| = 1$ realization of class DIII admits no relevant or marginal
time-reversal symmetric perturbations.  The $|\nu| = 2$ realization of class DIII is the same
as the minimal $|\nu| = 1$ version of AIII: a single complex Dirac fermion,
perturbed by an abelian $O(2)$ vector potential. 

We use the conformal embedding rules to obtain the effective
conformal field theories relevant for computing disorder-averaged
correlation functions in the non-interacting system. 
These are the level $k$ subalgebras in Eq.~(\ref{Embed}).
In the many-valley $k \gg 1$ limit for each TSC class, the non-abelian bosonization
of the corresponding conformal theory becomes weakly coupled.
This has exactly the same form as the non-linear sigma model
description of Anderson (de)localization expected in the appropriate symmetry class,\cite{Loc}
except that it is augmented by a Wess-Zumino-Novikov-Witten term. 
The symmetry of the target manifold is $G \times G/G$, with
$G \in \{\mathrm{Sp}(2 n), \mathrm{U}(n), \mathrm{O}(n)\}$ for class CI, AIII, and DIII, respectively;
$n$ is the number of replicas.
For classes AIII and DIII (CI), the level $k$ of the Wess-Zumino-Novikov-Witten 
model is (half of) the bulk winding number modulus.
This is also similar to the usual sigma model description of quantum
diffusion, in that the coupling is proportional to the inverse universal
Landauer conductance of the corresponding Dirac theory.

Unlike other applications of non-abelian bosonization at higher
levels $k > 1$ (such as multicritical spin chains),\cite{Affleck} 
perturbations that break the global 
$G \times G$ invariance of the sigma model target manifold are
forbidden. This is because the space of fluctuations (target manifold symmetry)
is determined by the 
random matrix class\cite{Loc} of the disordered Dirac Hamiltonian. 
To incorporate interactions, we need only interpret the
target manifold as a product of replica times Matsubara frequency spaces.
The allowed interactions can be inferred from the Dirac theory
via the conformal embedding map, or written down directly
in the sigma model based upon general considerations.
We emphasize that the structure of the sigma model emerges 
directly from the conformal embedding 
and from a symmetry analysis of the 
(2+1)-D imaginary time path integral for the surface Majorana band.\cite{Foster08,WZNW-P3}
We make no use of the  
self-consistent Born approximation or gradient expansion
employed in diffusive metals.\cite{AltlandSimons,SUSY,Altland02}
We therefore obtain an interacting (Finkel'stein)\cite{Finkelstein,BK}
non-linear sigma model without appealing to ideas of Fermi
liquid theory. 

The Majorana surface quasiparticles do not carry well-defined electric charge,
but can conduct energy as well as spin, if the latter is conserved. 
In the absence of interactions, the zero temperature (Landauer) 
spin conductance in classes CI and AIII is unmodified by disorder,
and assumes a finite universal value.\cite{Ludwig94,Tsvelik95,Ostrovsky06}
We argue in Sec.~\ref{Sec: TopProtCFT} that the same result applies to the 
thermal conductance (divided by the temperature) in class DIII.
Numerical evidence\cite{Hatsugai97,Ryu01,YZC-P1} suggests that the topology protects
both wavefunction delocalization and strict conformal invariance
in classes CI and AIII.

In a separate work,\cite{WZNW-P3} we establish that interaction-mediated 
Altshuler-Aronov corrections to the spin or thermal conductance are suppressed 
in the conformal limit for all three TSC classes.
This result obtains from the sigma models and is valid
to lowest order in $1/k$, but is non-perturbative in the interactions for 
classes CI and AIII.
In Ref.~\onlinecite{WZNW-P3}, working directly with the disordered Majorana surface 
theory, we also prove that the lowest order Hartree and Fock corrections
to the spin conductance vanish exactly for classes CI and AIII,
in every fixed disorder realization.
These results imply that the sigma model conductance parameter is pinned
to the universal value at the conformal point in classes CI and AIII. 
Determining the stability of the 
phase then reduces to analyzing the RG equations for the interactions.
We will conjecture that the same applies to class DIII, although
the latter is stable for weak enough interactions regardless. 
The focus of the present paper is this stability analysis, 
and we also provide the technical details of our conformal field theory calculations.  

The outline of this paper is as follows.
In Sec.~\ref{Sec: Results}, we present our results. 
We first transcribe low-energy field theories of TSC surface states in each class. 
We summarize the critical properties of non-interacting surface
states in the presence of disorder. 
We then turn to the combined effects of disorder and interactions.
We review previous results\cite{Foster12-B} in class CI, and present new results
for AIII and DIII.
We enumerate the interaction channels and scaling dimensions of the corresponding coupling strengths.
We write down the one-loop RG equations for the corresponding WZNW-FNLsMs, valid in the large surface valley limit. 
Finally we construct global phase diagrams as a function of the interaction coupling strengths.

In Sec.~\ref{Sec: CFT}, we derive the conformal field theory results discussed
in Sec.~\ref{Sec: Results}. The derivations of the one-loop RG equations for the WZNW-FNLsMs 
appear elsewhere.\cite{WZNW-P3}
We conclude with open questions in Sec.~\ref{Sec: Conc}.


\section{Models and results \label{Sec: Results}}

\subsection{Majorana surface bands for 3D topological superconductors}

Different from a 3D $\mathbb{Z}_2$ topological insulator,
the signature of a bulk topological superconductor (TSC) is the presence
of any number of gapless, delocalized quasiparticle bands
wrapping the sample surface. The absolute value of the 
bulk winding number\cite{SRFL08} $|\nu|$ is equal
to the number of independent species (or valleys) of surface
quasiparticle bands. 

In the periodic table\cite{SRFL08,Kitaev09,SRFL10}, 
there are three classes of topological superconductors in 3D denoted
CI, AIII, and DIII. All are protected by time-reversal symmetry.
The classes are distinguished by the degree of spin SU(2) symmetry 
enjoyed by the bulk and surface.
Class CI requires full spin SU(2) invariance, even in the presence
of disorder. A 3D class CI topological superconductor would require
spin singlet pairing and a fully-gapped bulk with negligible spin-orbit coupling.
Class AIII is distinguished by spin U(1) invariance, as could
be realized in a spin-triplet p-wave\cite{Foster08} TSC. Class 
DIII assumes no spin symmetry, e.g.\ a TSC with strong spin-orbit coupling. 
Cu${}_x$Bi$_2$Se$_{3}$ has been proposed as a possible realization of 
class DIII.\cite{Fu10,Hsieh12,ExpCuBiSe1,ExpCuBiSe2,ExpCuBiSe3,ExpCuBiSe4,ExpCuBiSe5}
The $B$ phase of ${}^3$He is a bulk topological superfluid in the same class.\cite{Volovik,SRFL08}

The form of the low-energy surface Dirac theory can be obtained 
in various ways. In Ref.~\onlinecite{SRFL08}, two general
arguments were given. One is a topological field theory construction,
while the second links the existence of a topological bulk in
classes CI, AIII, and DIII to the ``extended'' random matrix 
classification\cite{BL02} of Dirac fermions in 2D.
A third way is to extract the surface states explicitly from
a microscopic lattice model.\cite{SRL09,Foster12-B} 
A derivation of this type for class AIII appears in
Ref.~\onlinecite{WZNW-P3}. The main output of this
procedure is to verify the form of time-reversal 
symmetry, Eqs.~(\ref{ChiralTRI}) and (\ref{ChiralTRIop}).

\subsubsection{Spin SU(2) symmetry: CI}

In class CI, surface quasiparticle bands appear in 
valley degenerate pairs.\cite{SRFL08,SRL09} 
For a bulk with winding number $\nu = \pm 2 k$, $k \in \{1,2,\ldots\}$,
the low-energy Dirac theory can always be written as\cite{Foster12-B}  
\begin{align}\label{hci}
	\hci
	=&
	\frac{1}{2 \pi}
	\int
	d^2 \vex{r} \,
	\psi^\dagger
	\left\{
	\sigb \cdot
	\left[
	i \Nabla
	-
	\vex{A}_i(\vex{r}) \, 
	\tk^i
	\right]
	\right\}
	\psi.
\end{align}
The complex Dirac field $\psi \rightarrow \psi_{\sigma,v,a}$ is a $4 n k$-component
spinor, with indices in pseudospin $\sigma \in \{1,2\}$, valley $v \in \{1,2,\ldots,2 k\}$, and
replica $a \in \{1,2,\ldots,n\}$ spaces.
In this case the pseudospin matrices $\sigh^{1,2,3}$
act on a combination of Nambu (particle-hole) and orbital degrees of freedom,
the precise interpretation of which requires a bulk microscopic model.\cite{SRL09,Foster12-B}
The vector operator $\sigb = \{\sigh^1,\sigh^2\}$.

Time-reversal symmetry is encoded in the chiral transformation\cite{BL02,SRFL08}
\begin{align}\label{ChiralTRI}
	\psi \rightarrow \sigh^3 (\psi^\dagger)^\T, \;\; i \rightarrow -i,
\end{align}
where $\T$ denotes the matrix transpose.
This transformation is antiunitary, and is equivalent to the condition 
\begin{align}\label{ChiralTRIop}
	-\hat{M}_S \, \hat{h} \, \hat{M}_S = \hat{h}, \;\; \hat{M}_S = \sigh^3
\end{align}
for the single particle Hamiltonian $\hat{h}$ in $\hci \equiv \int \psi^\dagger \hat{h} \psi$. 

That time-reversal invariance can be represented as a chiral symmetry is
true for any Bogoliubov-de Gennes quasiparticle Hamiltonian
(gapped or gapless, topological or trivial), and is a mere consequence of the particle-hole redundancy.\cite{Foster08}
For 2D Dirac quasiparticles in classes CI, AIII, and DIII, only
two unitarily inequivalent choices are possible for the conjugating matrix $\hat{M}_S$.\cite{BL02} 
In the topological case [Eqs.~(\ref{ChiralTRI}) and (\ref{ChiralTRIop})], 
this restricts allowed perturbations to be off-diagonal in pseudospin space. 
As a consequence, only vector potentials are allowed.\cite{SRFL08} 
By contrast, for 2D Dirac quasiparticles in topologically trivial class CI superconductors
such as the $d$-wave cuprates, time-reversal symmetric perturbations include
Dirac mass and scalar potentials terms.\cite{Altland02} In the latter case, 
$\hat{M}_S = \sigh^3 \hat{N}$, where $\hat{N}$ is a diagonal matrix in valley space
with $p$ ($q$) elements equal to $+1$ ($-1$), $p,q \geq 1$, and 
$p + q = 2 k$ ($2k$ is the number of valleys).\cite{BL02}
The appearance of pseudomagnetic fields in a time-reversal invariant
quasiparticle Hamiltonian is not a feature unique to 
topological superconductors. Rather, the special feature
is the exclusion of other perturbations.

The vector potential $\vex{A}_i(\vex{r}) \, \tk^i$ in Eq.~(\ref{hci}) encodes generic
time-reversal symmetric surface potentials. These could include impurity charges
or external electric fields, surface deformations of the bulk pairing strength,
as well as edge, corner, dislocation or disclination potentials. 
The vector potential is non-abelian, coupling the $2 k$ valleys through generators 
$\tk^i \rightarrow (\tk^i)_{v}{}^{v'}$ of the valley symmetry group Sp$(2 k)$. 
Here the lower (upper) index transforms in the fundamental (conjugate) representation.
These are equivalent through raising or lowering by the symplectic ``metric''
$(\kah^2)^{v,v'} = (\kah^2)_{v,v'}$, which is a $2 k \times 2 k$ antisymmetric
block Pauli matrix.

Spin SU(2) symmetry is hidden in Eq.~(\ref{hci}); the U(1) charge
of $\psi$ is the spin-projection along the z-axis (say).
A spin rotation by $\pi$ around the x-axis appears as the 
unitary particle-hole ($P$) transformation
\begin{align}\label{Sx}
	\psi \rightarrow \hat{M}_P (\psi^\dagger)^\T, \;\;
	\hat{M}_P = i \sigh^1 \kah^2. 
\end{align}
Because $\hat{M}_P^\T = - \hat{M}_P$, we have $P^2 = -1$ (class CI).\cite{SRFL08,Kitaev09,SRFL10}
Equation (\ref{Sx}) restricts the non-abelian vector potential in 
Eq.~(\ref{hci}) to Sp$(2 k)$ generators acting on valley space, 
since
\[
	-\kah^2 (\tk^i)^\T \kah^2 = \tk^i.
\]
The abelian Dirac 3-current encodes the z-spin density and associated spin current,
\begin{align}\label{AbelCurr}
	\psi^\dagger \psi(\vex{r}) = 2 S^z(\vex{r}),
	\quad
	\psi^\dagger \sigb \psi(\vex{r}) = 2 \vex{J}^z(\vex{r}). 
\end{align}
(Positive and negative spin densities are possible because the Dirac
field has particle and antiparticle excitations). 
The z-spin density (current) is odd (even) under time-reversal.
The x- and y-spin 3-currents are anomalous in the $\psi$ language,
i.e.\ involve terms such as $\psi^\T \sigh^1 \kah^2 \psi$ and $\psi^\dagger \sigh^1 \kah^2 (\psi^\dagger)^\T$.

A manifestly covariant formulation obtains by defining 
\begin{align}\label{CIDecomp}
\begin{aligned}
	\left\{\lf_{\uparrow,v,a},\lf_{\downarrow,v,a}\right\}
	\equiv&
	\left\{
	\psi_{1,v,a},\,
	\psi_{2,a}^{\dagger \, v'} (\kah^2)_{v',v} 
	\right\},
	\\
	\left\{\rt_{\uparrow,v,a},\rt_{\downarrow,v,a}\right\} 
	\equiv&
	\left\{
	\psi_{2,v,a},\,
	\psi_{1,a}^{\dagger \, v'} (\kah^2)_{v',v} 
	\right\},
	\\
	\LF \equiv&\, \lf^\T i \muh^2 \kah^2 \rightarrow \LF_a^{s,v},
	\\
	\RT \equiv&\, \rt^\T i \muh^2 \kah^2 \rightarrow \RT_a^{s,v}. 
\end{aligned}
\end{align}
The fields $\mathcal{L} \rightarrow \mathcal{L}_{s,v,a}$ and 
$\mathcal{R}_{s,v,a}$ transform in the fundamental representations of
the spin SU(2) and valley Sp$(2 k)$ 
symmetry groups. Here  $s \in \{\uparrow,\downarrow\}$ is the spin index. 
$\LF^{s,v}_a$ and $\RT_a^{s,v}$ transform in the conjugate representations;
$\hat{s}^2$ is the antisymmetric $2 \times 2$ Pauli matrix in spin space.
All four fields $\{\lf_a,\rt_a,\LF_a,\RT_a\}$ transform under the fundamental
representation of SO$(n)$ in replica space.
Introducing complex coordinates $\{z,\bar{z}\} = x \pm i y$ and defining 
$\{\partial,\bar{\partial}\} \equiv \frac{1}{2}(\partial_x \mp i \partial_y)$,
Eq.~(\ref{hci}) can be rewritten as 
\bsub\label{hciCFT}
\begin{align}
	\hci =&\, \hcio + \hcit,
	\\
	\label{hcioDef}
	\hcio
	=& \,
	-
	\frac{i}{2 \pi}
	\int d^2\vex{r}
	\left[
	\LF \, \bar{\partial} \, \lf
	+ 
	\RT \, \partial \, \rt
	\right],
	\\
	\label{hcitDef}
	\hcit
	=&
	\frac{1}{2 \pi}
	\int d^2\vex{r}
	\left[
	J_{\kappa}^i \bar{A}_i + \bar{J}^i_{\kappa} A_i
	\right].
\end{align}
\esub
In the last equation, 
\begin{align}\label{HoloVecs}
	\{A_i,\bar{A}_i\} = i (A^x_i \mp i A^y_i),
\end{align}
and
we have introduced the Sp$(2k)$ valley current operators
\begin{align}
\begin{aligned}
	J^i_{\kap}(z)
	\equiv&\,
	- 
	{\textstyle{\frac{i}{2}}} \LF \, \tk^i \, \lf
	=
	{\textstyle{\frac{1}{2}}} 
	\lf_{s,v,a} 
	\, 
	(\hat{s}^2)^{s,s'} (\kah^2 \tk^i)^{v,v'}
	\,
	\lf_{s',v',a},
	\\
	\bar{J}^i_{\kap}(\bar{z})
	\equiv&\,
	- 
	{\textstyle{\frac{i}{2}}} \RT \, \tk^i \, \rt
	=
	{\textstyle{\frac{1}{2}}} 
	\rt_{s,v,a} 
	\, 
	(\hat{s}^2)^{s,s'} (\kah^2 \tk^i)^{v,v'}
	\,
	\rt_{s',v',a}.
\end{aligned}
\end{align}
The current $J^i_{\kap}$ [$\bar{J}^i_{\kap}$] generates
transformations in the holomorphic $\mathcal{L}(z)$
[antiholomorphic $\mathcal{R}(\bar{z})$] sector of the theory.
Here and throughout the rest of this article,
doubly-repeated indices are summed unless otherwise indicated. 
Using Wick's theorem, one can check that the valley currents satisfy the 
Sp$(2 k)_n$ Kac-Moody algebra:
\begin{align}\label{CIValleyKM}
	J^i_{\kap}(z) 
	\,
	J^j_{\kap}(w)
	\sim&\,
	\frac{n \delta^{i j}}{(z - w)^2}
	+
	\frac{i f_{\kap}^{i j k}}{(z - w)}
	J^k_{\kap}(w),
\end{align}
where 
\begin{align}\label{StructsSp}
	\protect{[}\tk^{i}, \tk^{j} \protect{]}
	=&\, 
	i f_{\kap}^{i j k} \, \tk^k.
\end{align}
The normalizations in Eqs.~(\ref{CIValleyKM}) and (\ref{StructsSp}) 
reflect the trace and Fierz identities enumerated in Table~\ref{FierzTable}.

\begin{table}
\caption{Fierz and trace identities
	\label{FierzTable} 
	}
\begin{ruledtabular}
\begin{tabular}{l @{} l @{} l}
Algebra & Fierz identity & trace normalization\\
\hline
\vspace{-6pt}\\
	Sp$(2k)$ 
	& 
	$\sum_i (\kah^2 \hat{t}^i_\kap)^{m n} (\hat{t}^i_\kap \kah^2)^{p q}$
	& 
	$\tr\left[\hat{t}^i_\kap \hat{t}^j_\kap\right] = \delta^{i j}$ 
	\\
	& 
	$\quad = {\ts{\frac{1}{2}}}\left(\delta^{m q} \delta^{p n} +  \delta^{m p} \delta^{q n}\right)$
	& 
	\\
	\vspace{-6pt}
	\\
	SU$(k)$ 
	& 
	$\sum_i (\hat{t}^i_\kap)_m{}^{n} (\hat{t}^i_\kap)_{p}{}^{q}$
	& 
	$\tr\left[\hat{t}^i_\kap \hat{t}^j_\kap\right] = \delta^{i j}$
	\\
	& 
	$\quad = \delta_{m}{}^{q} \delta_{p}{}^n - {\ts{\frac{1}{k}}} \delta_{m}{}^{n} \delta_{p}{}^{q}$
	& 
	\\
	\vspace{-6pt}
	\\
	SO$(k)$
	& 
	$\sum_i (\hat{t}^i_\kap)^{m n} (\hat{t}^i_\kap)^{p q} $
	& 
	$\tr\left[\hat{t}^i_\kap \hat{t}^j_\kap\right] = 2 \, \delta^{i j}$
	\\
	& 
	$\quad = \delta^{m q} \delta^{p n} - \delta^{m p} \delta^{q n}$
	& 
	\\
	\vspace{-6pt}
	\\
\end{tabular}
\end{ruledtabular}
\end{table}

The fields $\lf_{s,v,a}$ and $\rt_{s,v,a}$ are each $4 n k$-component spinors. 
Eq.~(\ref{hcioDef}) is invariant under independent left and right
SO$(4 n k)$ transformations acting on the combined 
spin $\otimes$ valley $\otimes$ replica spaces: 
\begin{align}
\begin{gathered}
	\lf \rightarrow \hat{O} \, \lf, \;\;
	\LF \rightarrow \LF \, \hat{O}^{-1},
	\\
	\rt \rightarrow \hat{\overline{O}} \, \rt, \;\;
	\RT \rightarrow \RT \, \hat{\overline{O}}^{-1},
	\\
	\hat{O}^{-1}
	=
	\sigh^2 \kah^2 \, \hat{O}^\T \, \sigh^2 \kah^2,
	\;\;
	\hat{\overline{O}}^{-1}
	=
	\sigh^2 \kah^2 \, \hat{\overline{O}}^\T \, \sigh^2 \kah^2.
\end{gathered}
\end{align}
We can replace the Hamiltonian with a (2+0)-D Grassmann path integral; then
the clean theory with $A_i = \bar{A}_i = 0$ is equivalent
to the SO$(4 n k)_1$ conformal field theory. 
The class CI embedding scheme in Eq.~(\ref{Embed}) implies 
that this theory can be decomposed into Sp$(2 k)_n$ valley and Sp$(2 n)_k$ 
spin $\otimes$ replica sectors.

\subsubsection{Spin U(1) symmetry: AIII}

The winding number is integer-valued for class AIII, so that
the surface quasiparticle action for a TSC can have any number 
of $k \in \{1,2,\ldots\}$ valleys.
Time-reversal symmetric surface perturbations are non-abelian and abelian 
vector potentials that couple to the valley and spin $U(1)$
currents, respectively. 
The minimal case of $k = 1$ is special, in that only
the abelian spin current can appear. 
The latter model has been extensively studied.\cite{Ludwig94,Chamon96,Castillo97,Hatsugai97,Ryu01,Carpentier01,Motrunich02,Horovitz02,Mudry03}
The non-abelian case was analyzed in Refs.~\onlinecite{Nersesyan94,Tsvelik95,Bernard95,Mudry96,Caux96,Ludwig00,Bhaseen01,Ostrovsky06}.

The Hamiltonian for an AIII TSC with winding number $k > 1$ is 
\begin{align}\label{hai}
	\hai
	=&
	\frac{1}{4 \pi}
	\int
	d^2 \vex{r} \,
	\psi^\dagger
	\left\{
	\sigb \cdot
	\left[
	i \Nabla
	-
	\vex{A}_i(\vex{r}) \, 
	\tk^i
	-
	\vex{A}(\vex{r})
	\right]
	\right\}
	\psi.
\end{align}
The complex field $\psi$ is interpreted the same way as in class CI, 
except that the valley index $v \in \{1,2,\ldots,k\}$.
The U(1) charge corresponds to the conserved z-spin density. 
The abelian vector potential $\vex{A}$ couples to the
corresponding current [Eq.~(\ref{AbelCurr})].
The non-abelian potential $\vex{A}_i$ couples
to SU$(k)$ generators $\tk^i$ acting on valley space. 

We define the pseudospin decomposition,
\begin{align}\label{AIIIDecomp}
\begin{aligned}
	\psi_{v,a} \equiv
	\begin{bmatrix}
	L_{v,a} \\
	R_{v,a}
	\end{bmatrix},
	\quad
	(\psi^\dagger)^{v,a} \equiv 
	\begin{bmatrix}
	R^{\dagger\, v,a} &
	L^{\dagger\, v,a}
	\end{bmatrix},
\end{aligned}
\end{align}
which leads to
\bsub\label{haiCFT}
\begin{align}
	\hai =&\, \haio + \hait,
	\\
	\label{haioDef}
	\haio
	=& 
	\,
	\frac{i}{2 \pi}
	\int d^2\vex{r}
	\left[
	L^\dagger \, \bar{\partial} \, L
	+ 
	R^\dagger \, \partial \, R
	\right],
	\\
	\label{haitDef}
	\hait
	=&
	\frac{1}{2 \pi}
	\int d^2\vex{r}
	\left[
	J_{\kappa}^i \bar{A}_i + \bar{J}^i_{\kappa} A_i + J \bar{A} + \bar{J} A
	\right].
\end{align}
\esub
In the last equation, the vector potential components
are defined as in Eq.~(\ref{HoloVecs}), while 
\begin{align}
\begin{aligned}
	J_{\kappa}^i(z)
	=&\,
	- 
	i 
	\gamma_\kappa
	\,
	L^{\dagger\, v,a} 
	(\tk^i)_{v}{}^{v'}
	L_{v',a},
	\\
	J(z)
	=&\,
	- 
	i
	\gamma_0
	\,
	L^{\dagger\, v,a} 
	L_{v,a}.
\end{aligned}
\end{align}
The constants $\gamma_{\kappa,0}$ are chosen so that
$J_{\kappa}^i(z)$ [$J(z)$] satisfies the canonical
SU$(k)_n$ [U(1)] Kac-Moody algebra. 

Eq.~(\ref{haioDef}) is invariant under U$(n k) \times$U$(n k)$ 
replica $\otimes$ valley $\otimes$ spin U(1) transformations,
and is equivalent to U$(n k)_1$. 
The AIII embedding scheme in Eq.~(\ref{Embed}) implies
that this can be decomposed into 
SU$(k)_n$ valley 
and
spin U(1) $\otimes$ replica 
SU$(n)_k$ 
sectors.\cite{Nersesyan94,Tsvelik95,Caux96}

\subsubsection{No spin symmetry: DIII}

In class DIII, the winding number $\nu \in \mathbb{Z}$ and
there are $k = |\nu|$ surface quasiparticle bands. 
Spin SU(2) invariance is completely destroyed,
as occurs in the presence of strong spin-orbit coupling.
Another way to realize class DIII surface states is 
to deposit strong spin-orbit scattering impurities
on the surface of an erstwhile class CI 
or AIII 
TSC. 

Because spin is not conserved, there is no natural U(1)
with which to define complex Dirac spinors for the
surface quasiparticle bands. Instead, we are forced
to work with a real Majorana spinor $\chi_{\sigma,v,a}$,
with pseudospin 
($\sim$ real spin)
index $\sigma \in \{1,2\}$, valley 
$v \in \{1,2,\ldots,k\}$, and
replica $a \in \{1,2,\ldots,n\}$.
The $k = 1$ case admits no relevant or marginal perturbations,
so (weak) disorder has a negligible effect. 
The $k = 2$ case is identical to AIII with $k = 1$,
a single complex Dirac fermion coupled to an abelian
vector potential. In this case the U(1) charge is the valley
polarization.

For $k \geq 3$ (three or more valleys), DIII admits
non-abelian intervalley scattering. 
The Hamiltonian is 
\begin{align}\label{hdi}
	\hdi
	=
	\frac{1}{4 \pi}
	\int
	d^2 \vex{r} \,
	\chi^\T
	\hat{M}_P
	\left\{
	\sigb \cdot
	\left[
	i \Nabla
	-
	\vex{A}_i(\vex{r}) \, 
	\tk^i
	\right]	
	\right\}
	\chi,
\end{align}
where 
\begin{align}
	\hat{M}_P = -i \sigh^1
\end{align}
is a particle-hole conjugation matrix. ($\hat{M}_P^\T = \hat{M}_P$, so that
$P^2 = +1$, class DIII.)
Time reversal ($T$) is encoded by the transformation
\begin{align}
	\chi \rightarrow \hat{M}_T \chi, \;\; i \rightarrow - i, \;\; \hat{M}_T = -i \sigh^2, 
\end{align}
which implies that $T^2 = -1$. 
The vector potential couples to valley space generators $\{\tk^i\}$ for 
SO$(k)$, since these must satisfy
\[
	-(\tk^i)^\T = \tk^i.
\]

With the decomposition
\begin{align}\label{DIIIDecomp}
	\chi_{v,a} \equiv 
	\begin{bmatrix}
	L_{v,a} \\
	R_{v,a}
	\end{bmatrix},
\end{align}
the Hamiltonian can be rewritten as 
\bsub\label{hdiCFT}
\begin{align}
	\hdi =&\, \hdio + \hdit,
	\\
	\label{hdioDef}
	\hdio
	=& 
	\,
	\frac{1}{2 \pi}
	\int d^2\vex{r}
	\left[
	L \, \bar{\partial} \, L
	+ 
	R \, \partial \, R
	\right],
	\\
	\label{hditDef}
	\hdit
	=&
	\frac{1}{2 \pi}
	\int d^2\vex{r}
	\left[
	J_{\kappa}^i \bar{A}_i + \bar{J}^i_{\kappa} A_i
	\right].
\end{align}
\esub

The valley currents $J_{\kappa}^i(z) = - i \gamma_k L^\T \tk^i L$ 
in Eq.~(\ref{hditDef}) generate an SO$(k)_n$ Kac-Moody algebra.
The free Hamiltonian in Eq.~(\ref{hdioDef}) is invariant
under SO$(n k) \times$ SO$(n k)$ replica $\otimes$ valley space
transformations, and is equivalent to $SO(n k)_1$. 
The DIII embedding in Eq.~(\ref{Embed}) implies that this
can be decomposed into SO$(k)_n$ valley and SO$(n)_k$ replica sectors.

\subsubsection{Spin and thermal quantum Hall effects: 
Dirac mass order parameters for imaginary surface state Cooper pairing 
\label{Sec: QHEmass}}

For each of the three classes of TSCs there
is a single valley-invariant Dirac surface mass operator:
\begin{align}\label{mass}
\begin{aligned}
	m(\vex{r}) =&\, \psi^\dagger \sigh^3 \psi,&& \;\; \textrm{CI and AIII [Eqs.~(\ref{hci}), (\ref{hai})]}
	\\
	m(\vex{r}) =&\, \chi^\T \hat{M}_P \, \sigh^3 \chi,&& \;\; \textrm{DIII [Eq.~(\ref{hdi})]}.
\end{aligned}
\end{align} 
In each case, this operator breaks time-reversal invariance and opens 
up a gap in the surface spectrum at the Dirac point, but preserves
spin SU$(2)$ [U$(1)$] symmetry in class CI (AIII). 

In fact, a Dirac mass $\langle m \rangle \neq 0$ induces a surface quantum Hall
effect, analogous to the ``half-integer'' quantum Hall phase at the surface of a 3D $\mathbb{Z}_2$
topological insulator with broken time-reversal symmetry.\cite{TISC,Ryu12,Stone12} 
In each case, the mass modifies the \emph{surface state} symmetry class.
Class CI shifts to class C; the gapped surface resides in a  
plateau of the SU$(2)$ invariant \emph{spin quantum Hall effect}.\cite{SpinQHE,SRL09}
This state is characterized by a quantized spin Hall conductance\cite{SpinQHE,Foster12-B} 
\begin{align}\label{SpinQHECond}
	\sigma^s_{x y} = \frac{1}{h}\left(\frac{\hbar}{2}\right)^2  k \sgn \langle m \rangle,
\end{align}
where $k = |\nu|/2$ is half the number of surface valleys, and
$\hbar/2$ is the ``spin charge.'' 
The spin Hall current could be induced on a closed TSC surface divided
into two complimentary domains $A$ and $B$, with $\langle m \rangle > 0$ ($\langle m \rangle < 0$)
in domain $A$ ($B$). 
For $k = 1$, the domains would be separated by a one-dimensional, 
two-channel edge state (corresponding to up and down spins). Applying a linearly varying
Zeeman field across domain $A$ (say) would induce a circulating spin edge current with
polarization equal to that of the external field.  
The physics is similar in class AIII, which shifts to class A. 
The gapped phase is a plateau of the ``ordinary'' quantum Hall effect,
with the conserved z-spin taking the place of electric charge.  

A topological superconductor in class DIII conserves only heat,
so a mass term induces a thermal quantum Hall state. Majorana
edge states separating spatial domains in different plateaux 
demonstrate a quantized thermal Hall conductivity\cite{ReadGreen2000,Ryu12,Stone12}
in the presence of a transverse temperature gradient. 

How can a non-zero Dirac mass be induced? 
For a TSC, it turns out that $\langle m \rangle$ can be viewed
as the order parameter amplitude for \emph{imaginary} Cooper pairing of
the surface Majorana quasiparticles.\cite{Foster12-B} 
In other words, due to attractive interactions the gapless 
surface quasiparticles can Cooper pair at a non-zero superfluid phase angle relative to
the bulk. This is necessary to open a gap and break
time-reversal symmetry; pairing of the surface at the same angle
as the bulk would appear as a vector potential order parameter,
which neither opens the gap nor localizes the surface quasiparticles. 
Indeed, a homogeneous time-reversal invariant pairing merely
shifts the surface Dirac points, and can be removed 
from the low energy theory by a gauge transformation. 

For a clean TSC surface, short-ranged interactions are
strongly irrelevant, and Cooper pairing is only possible 
for relatively strong interactions. 
However, disorder can \emph{enhance} interaction effects,
due to the strong spatial inhomogeneities induced 
in delocalized wavefunctions by quantum interference.\cite{Foster12-B,IQHP,Feigelman07,Feigelman10}
In class CI, it turns out interactions have a strong 
effect for arbitrarily weak disorder, due to the
enhancement of interaction matrix elements by 
multifractal fluctuations in the local density of states 
(Ref.~\onlinecite{Foster12-B}; see also Sec.~\ref{CIResults}, below). 
For any winding number $|\nu| = 2 k \neq 0$ in class CI, 
time-reversal breaks spontaneously at the surface.
The dominant interaction channel favors $\langle m \rangle \neq 0$.

One can define other types of Dirac mass operators that
do not preserve valley symmetry. These can serve as
order parameters for states with valley space symmetry
that is broken \emph{on average} (after summing over
disorder realizations). The non-invariant mass operators 
also break time-reversal symmetry. Tuning the amplitudes of 
these relative to $\langle m \rangle$, one can drive the surface 
across plateau transitions into different surface Hall states.  
Mass operators of this type in class CI are discussed
in Appendix~\ref{App: NonInvtMass}.

Dirac mass operators appear again in Sec.~\ref{IntStabResults}, wherein
we discuss surface interaction channels and instabilities.

\subsection{Critical delocalization: non-interacting properties \label{Sec: DirtyCFTs}}

In this section we describe the influence of disorder on non-interacting 
topological superconductor surface states. 
The objects of study are the density of states exponents and multifractal spectrum
defined below via Eqs.~(\ref{DoSExpDef}) and (\ref{tauqDef}), respectively. 
Exact conformal field theory results for all three classes appear in 
Eqs.~(\ref{DoSExp}) and (\ref{MFCDim}), most of which were previously known.
In subsection~\ref{Sec: TopProtCFT}, we address the stability of these 
effective theories in the absence of interactions, and discuss evidence 
that topological protection extends to strict conformal invariance. 

We briefly summarize the logic of the approach. Technical details
are relegated to Sec.~\ref{Sec: CFT}.
To treat the influence of disorder, we can write a (2+0)-D
Grassmann path integral with the action given by Eqs.~(\ref{hciCFT}), 
(\ref{haiCFT}), or (\ref{hdiCFT}) for the appropriate symmetry class.
One can view this as the zero Matsubara frequency component of an 
imaginary time formulation. Without interactions, different
frequencies decouple. The strong correlations between any two (nearly) delocalized
wavefunctions as a function of their energy separation is known as Chalker 
scaling,\cite{Chalker88,Chalker90,Cuevas07,YZC-P1}
and is a universal characteristic of disorder-mediated quantum interference. 
Chalker scaling implies that low-energy properties can be extracted from the 
dimensions of local operators in the effective zero energy theory. 
These include the scaling behavior of the density of states with energy\cite{Ludwig94,Nersesyan94}
and the static enhancement or suppression of interaction matrix elements due 
to the disorder.\cite{IQHP,Feigelman07,Feigelman10,Foster12-B} 
We use replicas to average over disorder configurations,\cite{AltlandSimons}
with the number of replicas $n \rightarrow 0$ at the end.

For any amount of non-abelian valley disorder on the surface
of a 3D topological superconductor, the description in terms
of the clean Dirac bandstructure is unstable in the sense of the 
renormalization group (RG).\cite{Nersesyan94} 
For example, consider Gaussian white noise correlated disorder,
\begin{align}\label{VecVarNA}
	\overline{A_i^\alpha(\vex{r}) \, A_j^\beta(\vex{r'})} 
	= 
	\lamN \,
	\delta_{i j} \,
	\delta^{\alpha \beta}
	\delta^{\pup{2}}(\vex{r} - \vex{r}'),
\end{align}
where the overline denotes an ensemble average over disorder realizations, 
and $\lamN > 0$ is the disorder variance. 
Using the Kac-Moody operator product expansion for the valley
currents as in Eq.~(\ref{CIValleyKM}), the one-loop renormalization group (RG)
equation is\cite{Cardy}
\begin{align}\label{NARG}
	\frac{d \lamN}{d l}
	=
	2 \pi g \lamN^2 
	+
	\ord{\lamN}^3.
\end{align}
Here $g$ is the dual Coxeter number (half the quadratic Casimir
in the adjoint representation) 
for the valley symmetry group\cite{CFT}
\begin{align}\label{Coxnums}
	\begin{aligned}
	g[\textrm{Sp}(2 k)] =&\, k + 1, \\
	g[\textrm{SU}(k)] =&\, k, \\
	g[\textrm{SO}(k)] =&\, k - 2.
	\end{aligned}
\end{align}
As discussed below Eq.~(\ref{Embed}), $k \geq 2$ ($k \geq 3$) for class AIII (DIII),
so that Eq.~(\ref{NARG}) implies that the effects of disorder grow stronger
as we renormalize towards longer length scales and lower energies. 

A diverging RG flow implies that the correct effective field theory is strongly
coupled relative to the free field (clean Dirac) fixed point. 
Note, however, that the non-abelian disorder couples \emph{only} to the valley 
Kac-Moody currents in each of Eqs.~(\ref{hcitDef}), (\ref{haitDef}), and 
(\ref{hditDef}). 

The conformal embeddings in Eq.~(\ref{Embed}) imply that
the clean theory in Eqs.~(\ref{hcioDef}), (\ref{haioDef}), and (\ref{hdioDef})
can be decomposed into the ``sum'' of the level $n$ valley algebra [$\equiv G(k)_n$], 
and the level $k$ replica or replica $\otimes$ spin algebra [$\equiv H(n)_k$]. 
What this means is that the holomorphic stress tensor $T(z)$ for the clean theory 
can be written as the sum\cite{CFT}
\begin{align}
	T(z) = T_n^{\pupsq{G(k)}}(z) + T_k^{\pupsq{H(n)}}(z),
\end{align} 
where 
\[
	T_{q}^{\pupsq{G}}(z) = \frac{1}{2(q + g)} : J^\alpha J^\alpha :
\]
is the Sugawara stress tensor for the level $q$ algebra $G_q$ 
($g$ is the dual Coxeter number). 
Roughly speaking (but see e.g.\ Ref.~\onlinecite{QCD}), the full operator
content of the free level one theory in each embedding of Eq.~(\ref{Embed})
can be decomposed into Kac-Moody currents of the $G(k)_n$ and $H(n)_k$ subalgebras,
or products of primary and/or descendant fields from these two sectors. 

Because the disorder is a relevant perturbation in the sense of 
Eq.~(\ref{NARG}), yet couples only to the valley algebra $G(k)_n$
in each symmetry class, the effective low-energy, long-wavelength theory 
of the disorder-averaged TSC surface state is the ``conformal
remnant'' $H(n)_k$. 
That is, the conformal field theory that one can use to
calculate disorder-averaged correlation functions is
Sp$(2n)_k$ in class CI,\cite{SRL09,Foster12-B} 
U$(n)_k$ in class AIII,\cite{Nersesyan94,Tsvelik95} and
SO$(n)_k$ in class DIII. 
An alternative scheme based upon supersymmetric 
embeddings\cite{Mudry96,LeClair08} gives identical results.

Note that the effective level-$k$ theories know nothing about
the microscopic character of the non-abelian 
impurity potential.
Indeed, the main effect of the RG flow in Eq.~(\ref{NARG})
is to ``gauge away'' the disorder in valley space, leaving behind
the replica algebra with which we perform subsequent calculations.
A similar phenomenon occurs in (1+1)-D QCD, wherein the
color space becomes massive and decouples from a conformally
invariant flavor remnant.\cite{QCD} As a result, the predictions for
disorder-averaged correlation functions are universal for
classes CI and DIII. Class AIII admits an additional disorder parameter $\lambda_A$
[Eq.~(\ref{lambdaADef}), below],
which encodes the strength of vector potential fluctuations 
that couple to the \emph{abelian} spin U(1) current. 
The predictions for class AIII are therefore one-parameter
functions of this (see below).

We stress that for non-abelian disorder, it is not necessary to assume
the Gaussian white noise correlations in Eq.~(\ref{VecVarNA}). 
In general, white noise is a suitable replacement for any disorder potential
that is sufficiently short-range correlated. (The criterion is that momentum
space correlation function must exist as $\vex{k} \rightarrow 0$). For
TSC surface states, Coulomb impurities should be well-screened by the bulk
superfluid, but other sources of long-range correlated disorder could
exist.\cite{Fedorenko12} Long-range correlations produce a more strongly 
divergent RG flow,\cite{Fedorenko12} but we expect the same universal description of
the surface states described above (with the exception of
long-range correlated abelian disorder in class AIII).

\subsubsection{Density of states and multifractal spectra}

The average density of states $\nu(\e)$ is critical for Dirac fermions
coupled to vector potential disorder.\cite{Ludwig94,Motrunich02,Horovitz02,Mudry03,Nersesyan94,Tsvelik95}
The scaling behavior with respect to the energy $\e$ is determined by 
\begin{align}\label{DoSExpDef}
	\nu(\e) \sim |\e|^{x_1/z}.
\end{align}
Here, all energies are measured relative to the surface state Dirac point,
taken to reside at $\e = 0$.
In Eq.~(\ref{DoSExpDef}),
$x_1$ is the scaling dimension of the disorder-averaged
local density of states (LDoS) operator $\equiv \overline{\nu(\e,\vex{r})}$,\cite{footnote--LDoSOP}
and $z$ is the dynamic critical exponent. 
These are related by
\begin{align}\label{zDoS}
	2 = z + x_1.
\end{align}
For the three classes of disordered TSCs, one finds that
\begin{align}\label{DoSExp}
\begin{array}{lllll}
	\text{Class} &\quad & |\nu| &\quad & x_1/z \\
	\hline
	\\
	\vspace{-20pt}
	\\
	\text{CI} & &  2 k & & {\displaystyle{\frac{1}{4 k + 3}}}
	\\
	\vspace{-8pt}
	\\
	\text{AIII} & &  k \geq 1 & & {\displaystyle{\frac{1 - k^2 \lambda_A}{2 k^2 - 1 + \lambda_A k^2}}}
	\\
	\vspace{-8pt}
	\\
	\text{DIII} & &  k \geq 3 & & {\displaystyle{- \frac{1}{2 k - 3}}}
\end{array}
\end{align}

All of the results in Eq.~(\ref{DoSExp}) were previously known.
The result for AIII with $\lambda_A = 0$ first appeared in Ref.~\onlinecite{Nersesyan94}.
Here, $\lambda_A$ is the strength of the abelian vector potential disorder in 
Eqs.~(\ref{hai}) and (\ref{haitDef}):
\begin{align}\label{lambdaADef}
	\overline{A^\alpha(\vex{r}) \, A^\beta(\vex{r'})} 
	= 
	\lambda_A \,
	\delta^{\alpha \beta}
	\delta^{\pup{2}}(\vex{r} - \vex{r}').
\end{align}	
The parameter $\lambda_A$ is strictly marginal in the conformal limit, and
can be thought of as an effective ``Luttinger parameter'' in the spin U(1)
sector.
For a single valley $k = 1$, the AIII result above reproduces that of Ref.~\onlinecite{Ludwig94},
which first considered abelian disorder. 
The full expression for $x_1/z$ was implicit in Ref.~\onlinecite{Mudry96}.
Results for CI and DIII first appeared in Ref.~\onlinecite{LeClair08}. 

For class CI, the average density of states always vanishes as $\e \rightarrow 0$,
but the power approaches zero as $k \rightarrow \infty$. 
The same is true for class AIII for $\lambda_A < 1/k^2$,
while divergent behavior obtains in the opposite case. 
The average density of states for DIII is always divergent at zero energy.
Similar behavior
was found in the class DIII non-linear sigma model without the Wess-Zumino-Novikov-Witten
term.\cite{Senthil00}
In this paper we only discuss the regime of weak abelian disorder $\lambda_A < 2 + \frac{1-k}{k^2}$,
so that we avoid the ``freezing'' transition that occurs in the density of states\cite{Motrunich02,Horovitz02,Mudry03,YZC-P1}
and multifractal spectrum\cite{Chamon96,Castillo97,Carpentier01,YZC-P1} at stronger disorder. 

The disorder-induced spatial fluctuations of the LDoS $\nu(\e,\vex{r})$ 
are encoded in the \emph{multifractal spectrum}\cite{Paladin87,MFCRev,Loc} $\tau(q)$. 
The $\tau(q)$ spectrum measures the sensitivity of extended wavefunctions to the sample 
boundary. A large $L\times L$ area of the surface is finely partitioned
into a grid of boxes of size $a \ll L$. One then defines the
box probability $\mu_n$ and inverse participation ratio $\mathcal{P}_{q}$
in terms of a typical wavefunction $\psi(\vex{r})$:
\begin{align}\label{Box Prob}
	\mu_n
	\equiv
	\frac{
	\intl{\mathcal{A}_n}
	d^{2}\vex{r} \,
	|\psi(\vex{r})|^2
	}
	{
	\intl{L^2}
	d^{2}\vex{r} \,
	|\psi(\vex{r})|^2
	},
	\;\;\;
	\mathcal{P}_{q}
	\equiv
	\sum_{n} \mu_n^q,
\end{align}
where $\mathcal{A}_n$ denotes the $n^{\text{th}}$ box.
The normalization is such that $\mathcal{P}_{1} = 1$.

For a critically delocalized eigenstate $\psi$ (such as that found at a mobility edge), 
\begin{align}\label{tauqDef}
	\mathcal{P}_{q} \sim \left(\frac{a}{L}\right)^{\tau(q)},
\end{align}
where the exponent $\tau(q)$ is self-averaging and typically universal.\cite{Chamon96}
The multifractal spectrum thus provides a unique fingerprint for 
spatial fluctuations in a particular symmetry class. 
In the field-theoretic description,
the $q$th moment of the disorder-averaged LDoS ($q \in \{1,2,3\ldots\}$) 
is associated to a particular composite operator $\mathcal{O}_q$, with scaling dimension $x_q$. 
The set of such dimensions determines the multifractal spectrum via
\begin{align}\label{tauqDefLDoS}
	\tau(q) = 2(q - 1) + x_q - q x_1.
\end{align}
By contrast, localized states are insensitive to the sample boundary
for sufficiently large $L$ and have $\tau(q) = 0$.

For topological superconductor surface states,
the multifractal spectrum $\tau(q)$ is exactly quadratic:
\begin{equation}\label{MFCSpec}
	\tau(q) 
	=
	(q - 1)
	\left(2 - \theta_k q\right),
\end{equation}
where
\begin{align}\label{MFCDim}
\begin{array}{lllll}
	\text{Class} &\quad & |\nu| &\quad & \;\;\; \theta_k \\
	\hline
	\\
	\vspace{-20pt}
	\\
	\text{CI} & &  2 k & & {\displaystyle{   \frac{1}{2(k+1)}    }}
	\\
	\vspace{-8pt}
	\\
	\text{AIII} & &  k \geq 1 & & {\displaystyle{ \frac{k-1}{k^2} + \lambda_A       }}
	\\
	\vspace{-8pt}
	\\
	\text{DIII} & &  k \geq 3 & & {\displaystyle{  \frac{1}{k - 2}      }}
\end{array}
\end{align}
The results for CI and AIII were first obtained in Refs.~\onlinecite{Foster12-B} and \onlinecite{Mudry96,Caux96},
respectively. The DIII result is new. 
The class AIII result for $k = 1$ is that expected for a single Dirac fermion
subject to abelian vector potential disorder.\cite{Ludwig94}
[In fact, Eq.~(\ref{MFCSpec}) holds only for $|q| \leq q_c$, with $q_c = \sqrt{2/\theta_k}$.
Outside of this region the spectrum is linear (``termination'').]\cite{Chamon96,Castillo97,Carpentier01,Foster09,Loc}

The associated scaling exponents $\{x_q\}$ satisfy the symmetry relation\cite{Sym06,Sym11,Sym13}
\begin{align}\label{SymEq}
	x_{q} = x_{q_* - q},
\end{align}
with $q_* = 2$ for class CI\cite{Sym11} and $q_* = 0$ for class DIII.\cite{Sym13}
See Ref.~\onlinecite{footnote--Sym}.

Wavefunction multifractality plays the major role in our discussion of interaction effects
and the stability of topological superconductor surface phases [Sec.~\ref{IntStab}, below].
In addition, multifractal spectra can be extracted from real space LDoS data 
(obtained e.g.\ via STM).\cite{Yazdani,Mirlin2loops,Foster12-A}

\subsubsection{Topological protection of conformal invariance \label{Sec: TopProtCFT}}

The exact results in Eqs.~(\ref{DoSExp}) and (\ref{MFCDim}) obtain 
for the conformally invariant level $k$ subalgebras in Eq.~(\ref{Embed}). 
We have argued that these emerge through ``fractionalization'' of the 
disordered Dirac theories in the low-energy, long-wavelength limit. 
The mechanism involves the conformal embedding schemes in Eq.~(\ref{Embed}).
However, the precise character
of the RG flow connecting the perturbed Dirac theory to the effective
level $k$ conformal field theory (CFT) is not clear. 
In principle, we should consider perturbations away from the conformal
fixed points, in order to determine their stability. 
Here we focus exclusively on the non-interacting theory; interactions
are treated in the next section.

One crucial difference between the level $k$ CFTs employed here,
as compared to other applications of non-abelian bosonization, 
is that perturbations which break \emph{global} $G \times G$ invariance
of the corresponding sigma model target manifold are disallowed. 
The action for an affine Lie algebra $G_k$ is\cite{CFT}
\begin{align}\label{SWZNW}
	S_{G}^{\pup{k}}(\lambda) 
	=
	\frac{1}{\lambda} 
	\intl{\partial B} \frac{d^2 \vex{r}}{8 \pi l_\phi} \, \tr\left(\partial_\mu \hat{Q}^\dagger \partial_\mu \hat{Q}\right) 
	+ 
	k \WZNW, 
\end{align}
where the Wess-Zumino-Novikov-Witten term is defined via
\begin{align}\label{WZNW}
	\WZNW 
	=
	-
	i
	&
	\intl{B}
	\frac{d^3 \vex{y}}{12 \pi \l_\phi}
	\epsilon^{\alpha \beta \gamma}
	\nonumber\\
	&\times\!
	\tr\!\left[
	\!
	\left(\hat{Q}^\dagger \partial_\alpha \hat{Q}\right)\!
	\left(\hat{Q}^\dagger \partial_\beta \hat{Q}\right)\!
	\left(\hat{Q}^\dagger \partial_\gamma \hat{Q}\right)\!
	\right].	
\end{align}
Here $B$ denotes a 3-manifold with 2D boundary $\partial B$.
The former (latter) can be taken as the interior volume (closed surface) of the TSC.\cite{Konig12} 
The parameter $l_\phi$ is the Dynkin index of the corresponding group.
The matrix field $\hat{Q}(\vex{r})$ takes values in the compact Lie group $G$. Eqs.~(\ref{SWZNW})
and (\ref{WZNW}) are manifestly invariant under the $G \times G$ transformation
\[
	\hat{Q}(\vex{r}) \rightarrow \hat{U}_L \, \hat{Q}(\vex{r}) \, \hat{U}_R,
\]
where $U_{L,R}$ are independent, unitary global transformations in $G$. 
At the conformal fixed point,\cite{footnote--PCMWZNW} the coupling strength $\lambda$ of the gradient
term in Eq.~(\ref{SWZNW}) is equal to $1/k$.

Typically, global $G \times G$ invariance is an emergent property\cite{Affleck} of a 2D conformal field theory,
not present in a parent microscopic model.  
By contrast, Eqs.~(\ref{hciCFT}), (\ref{haiCFT}), and (\ref{hdiCFT}) are invariant under 
$H(n) \otimes H(n)$ left and right (holomorphic and antiholomorphic) group transformations, where 
$H(n) \in \{\text{Sp}(2n),\text{U}(n),\text{O}(n)\}$ for classes CI, AIII, and DIII, respectively.
This result is more general than the conformal invariance itself. It applies to the 
replicated Grassmann path integration encoding correlation functions of any $d$-dimensional 
single particle Hamiltonian in the associated symmetry class.\cite{Loc,Foster08,WZNW-P3}
Thus we cannot add ``mass'' terms such as $\tr(\hat{Q} + \hat{Q}^\dagger)$ or 
$\tr(\hat{Q})\tr(\hat{Q}^\dagger)$ to Eq.~(\ref{SWZNW}). 

The CFT is distinguished by \emph{local} $G \times G$ invariance,
which is not a property of the Dirac theories in 
Eqs.~(\ref{hci}), (\ref{hai}), and (\ref{hdi}) for non-zero disorder.
Tuning the gradient coupling $\lambda$ in Eq.~(\ref{SWZNW}) away from 
$1/k$ preserves global but not local invariance. Perturbing
around the CFT, the lowest order RG equation for the deformation
\[
	\delta\lambda \equiv \lambda - \frac{1}{k}
\]
is given by
\[
	\frac{d \ln \delta\lambda}{d l} = -\frac{2 g}{k + g} + \ord{\delta \lambda},
\]
where $k$ is the level and $g$ is the dual Coxeter number. For the level $k$ algebras
in Eq.~(\ref{Embed}) corresponding to the TSC classes, in the replica $n \rightarrow 0$ 
limit we have
\bsub\label{WZNWLagFlows}
\begin{align}
	& \text{CI} &&[\text{Sp}(2n)_k]: && \frac{d \ln \delta\lambda}{d l} = -\frac{2}{(k + 1)}, \label{CIWZNWFlow}
	\\
	& \text{AIII} &&[\text{SU}(n)_k]: && \frac{d \ln \delta\lambda}{d l} = 0, \label{AIIIWZNWFlow}
	\\
	& \text{DIII} &&[\text{SO}(n)_k]: && \frac{d \ln \delta\lambda}{d l} = \frac{4}{(k - 2)}, \label{DIIIWZNWFlow}
\end{align}
\esub
up to corrections of order $\delta \lambda$.
In the large-$k$ limit, the sigma model description with $\lambda \sim 1/k$ in Eq.~(\ref{SWZNW}) becomes
weakly coupled. Perturbing around the Gaussian fixed point\cite{Witten84} ($ \lambda = 0$) gives the complimentary 
flow equations
\bsub\label{WittenFlows}
\begin{align}
	& \text{CI} &&[\text{Sp}(2n)_k]: && \frac{d \lambda}{d l} = \lambda^2 \left[1 - (k \lambda)^2\right], \label{CIGFlow}
	\\
	& \text{AIII} &&[\text{SU}(n)_k]: && \frac{d \lambda}{d l} = 0, \label{AIIIGFlow}
	\\
	& \text{DIII} &&[\text{SO}(n)_k]: && \frac{d \lambda}{d l} = -2 \lambda^2 \left[1 - (k \lambda)^2\right], \label{DIIIGFlow}
\end{align}
\esub
up to corrections of order $\lambda^3 \sim 1/k^3$. 
Linearizing about the non-trivial fixed point
in each case,
Eqs.~(\ref{WittenFlows}) are consistent with
(\ref{WZNWLagFlows}) to lowest order in $1/k$.

For class AIII, we must also consider the renormalization of the $U(1)$ disorder strength $\lambda_A$.
This is given by\cite{GLL}
\begin{align}\label{GLLlambdaAFlow}
	\text{AIII: }\quad \frac{d \lambda_A}{d l} = \lambda^2 \left[1 - (k \lambda)^2\right]. 
\end{align}
We derive Eq.~(\ref{GLLlambdaAFlow}) in Appendix~\ref{App: lambdaAFlow}.

We consider each TSC class in turn.
Eqs.~(\ref{CIWZNWFlow}) and (\ref{CIGFlow}) imply that the class CI CFT is an attractive fixed point,
in the absence of interactions. We therefore expect the disordered Dirac theory [Eq.~(\ref{hci})]
to flow into the Sp$(2n)_k$ fixed point in the long wavelength limit.\cite{Foster12-B}
By contrast, the class DIII CFT is \emph{unstable} according to Eqs.~(\ref{DIIIWZNWFlow}) and
(\ref{DIIIGFlow}) (recall that $k \geq 3$ for class DIII).
Class DIII is known to possess a stable metallic phase in 2D.\cite{Senthil00}

Class AIII presents a particularly peculiar case: the parameter $\lambda$ is marginal [Eqs.~(\ref{AIIIWZNWFlow}) and (\ref{AIIIGFlow})],
but the abelian disorder strength $\lambda_A$ flows to strong coupling for $\lambda \neq 1/k$ [Eq.~(\ref{GLLlambdaAFlow})].
Although we have obtained our results by perturbing around the CFT or the Gaussian fixed point, 
the AIII flow equations (\ref{AIIIGFlow}) and (\ref{GLLlambdaAFlow}) are in fact exact,\cite{GLL} i.e.\ valid to all orders in $\lambda$ and $\lambda_A$. 
Clearly $\lambda > 1/k$ is unphysical, as Eq.~(\ref{GLLlambdaAFlow}) sends $\lambda_A$ (the positive-definite
variance of a disorder potential) towards negative values. The case of $\lambda < 1/k$ leads
to so-called ``Gade'' scaling,\cite{Gade} wherein the density of states exhibits a strong (yet integrable) divergence,
and the wavefunctions near zero energy are driven to strong multifractality.\cite{Gade,Motrunich02,Mudry03}
This type of behavior occurs for non-topological 2D class AIII systems,\cite{Gade,Motrunich02,Mudry03}
which can be localized for sufficiently strong disorder.\cite{Konig12} These include nodal p-wave
superconductors, wherein $\lambda_A$ characterizes quenched orientational fluctuations in the Cooper pair
wavefunction.\cite{Foster08} It is known that the runaway flow $\lambda_A \rightarrow \infty$
strongly enhances interaction effects. The result is interaction-stabilized Anderson localization.\cite{Foster06,Foster08} 

One is tempted to conclude that the Wess-Zumino-Novikov-Witten conformal field theories
do not describe the low-energy physics of classes AIII and DIII (even in the absence of interactions), 
because the RG flow appears to require fine-tuning to avoid the above-described instabilities.
However, there is good evidence to the contrary. First, one can show that the Landauer spin
conductance in classes CI and AIII for the disordered Dirac theories in Eqs.~(\ref{hci}) and (\ref{hai})
is universal, independent of the disorder strength.\cite{Ludwig94,Tsvelik95,Ostrovsky06}
For a class CI or AIII TSC with winding number $\nu$, one has 
\begin{align}\label{SpinLandCond}
	\sigma_{xx}^s = \frac{|\nu|}{\pi h}\left(\frac{\hbar}{2}\right)^2.
\end{align}
[This is the analog of the Landauer conductance for $|\nu|$ species of 
massless Dirac electrons doped to the Dirac point. In Eq.~(\ref{SpinLandCond}),
$\hbar/2$ replaces the electric charge $e$.]
For class DIII, one can artificially double the theory to introduce a fictitious U$(1)$ charge.\cite{Senthil00} 
Then the arguments in Ref.~\onlinecite{Ostrovsky06} imply that the associated charge conductance is 
universal, and the thermal conductivity obtains via the Wiedemann-Franz relation. 
The DIII Majorana system carries half the value obtained in the doubled system,\cite{Senthil00} 
leading to the universal result
\begin{align}\label{ThermalCond}
	\kappa = \frac{k \pi^2}{6} \frac{k_B^2}{\pi h} T
\end{align}
for $k = |\nu|$ valleys 
($\nu$ denotes the bulk winding number). Here we ignore the effects of irrelevant perturbations
such as the random modulation of the Fermi velocity.\cite{Nakai14}

The Landauer conductance is a universal coefficient times the parameter $k$ (proportional to the number of valleys). 
Except for the Wess-Zumino-Novikov-Witten term, Eq.~(\ref{SWZNW}) has the same structure as the replicated 
non-linear sigma model for quantum diffusion in the appropriate symmetry class, which can arise
from any 2D microscopic theory with the appropriate combination of time-reversal, particle-hole, and chiral symmetries.\cite{Loc,SRFL08} 
Then the coupling $\lambda$ is proportional to the dimensionless resistance, 
and should be pegged to the universal value $1/k$. 

Further evidence that $\lambda = 1/k$ is protected comes from considering interaction corrections
to the conductance. We have found that the Altshuler-Aronov corrections to $\lambda$ vanish
for $\lambda = 1/k$
in all three classes,
as discussed in Ref.~\onlinecite{WZNW-P3} and reviewed in Sec.~\ref{IntStabResults}, below.
This result is valid to all orders in one of the interactions 
(for classes CI and AIII),
but is perturbative in 
$\lambda$. 
Second, we have also extended the argument of Ref.~\onlinecite{Ostrovsky06} to the Hartree and
Fock spin conductance corrections in the Dirac language for classes CI and AIII. 
We find that these vanish exactly\cite{WZNW-P3} in every realization of the disorder.
The consistency of these two different approaches requires that $\lambda = 1/k$ in classes CI and AIII.

Finally, in a separate work\cite{YZC-P1} we have computed the density of states exponent $x_1/z$ [Eq.~(\ref{DoSExpDef})] and 
the multifractal spectrum $\tau(q)$ [Eq.~(\ref{MFCSpec})] numerically for a class AIII TSC
Dirac Hamiltonian with two valleys ($k = 2$). Our results\cite{YZC-P1} are consistent 
with the CFT, Eqs.~(\ref{DoSExp}) and (\ref{MFCDim}). 

We therefore expect that the strict conformal invariance of the non-interacting 
TSC surface states is topologically protected. This implies that a disordered,
non-interacting class AIII surface will not exhibit Gade scaling.\cite{Gade,Motrunich02,Mudry03}
We expect that the thermal metal phase described in Ref.~\onlinecite{Senthil00} 
is not realized at the surface of a class DIII TSC.

\subsection{Interactions: Methods, multifractality, and Chalker scaling \label{IntStab}}

In what follows, we discuss the effects of interactions on disordered TSC surface 
states from multiple perspectives.
First, we enumerate all short-ranged four-fermion interactions consistent with
bulk symmetries (e.g., time-reversal invariance).
We do not consider long-ranged Coulomb interactions, since these should be 
screened by the bulk superfluid.
Using the conformal field theories (CFTs) described in Sec.~\ref{Sec: DirtyCFTs}, we
have computed the exact scaling dimensions of each to determine their relevance or irrelevance
as perturbations to the non-interacting fixed points. We report these results
and discuss their implications. We also give the full 1-loop RG equations for
the Wess-Zumino-Novikov-Witten Finkel'stein non-linear sigma models (WZNW-FNLsMs) 
in classes CI and AIII.
These are effective field theories obtained from non-abelian
bosonization of the associated CFTs. The one-loop RG is useful in the
$k \gg 1$ large winding number/many-valley limit. These equations can include
the effects of interactions to all orders, but are valid only to the lowest
order in $1/k$. We discuss the stability of TSC surface states by combining 
results and insights gained from the CFT and WZNW-FNLsM calculations.

The multifractal\cite{Paladin87} character of critically delocalized wavefunctions\cite{MFCRev,Loc} 
plays a crucial role in modulating interaction effects for TSC surface states. 
This is because generic short-ranged interactions are strongly irrelevant in the absence of disorder.
Multifractality means that individual wavefunctions are highly inhomogeneous in space,
with large accumulations in rare regions.\cite{MFCRev,Loc}  
Yet the remarkable phenomenon of Chalker scaling\cite{Chalker88,Chalker90,Cuevas07}
implies that such accumulations in single particle wavefunctions with nearby energy eigenvalues 
are strongly correlated. In other words, wavefunctions close in energy exhibit overlapping peaks
in position space. 
This can be understood by defining the energy-split inverse participation ratio\cite{Cuevas07,YZC-P1}
[C.f.\ Eq.~(\ref{Box Prob})]
\begin{align}\label{P2}
	\mathcal{P}_2(\e,L) \equiv \int d^2\vex{r} \, |\psi_0(\vex{r})|^2 |\psi_\e(\vex{r})|^2,
\end{align}
where $\psi_\e(\vex{r})$ is a representative wavefunction with eigenenergy $\e$.
For critically delocalized states such as those near zero energy at the surface of a dirty
topological superconductor, one has the asymptotic behaviors
\bsub\label{SplitIPR}
\begin{align}
	P_2(\e \rightarrow 0,L) \sim&\, \frac{1}{L^{\tau(2)}}, \label{IPRZE}
	\\
	P_2(\e, L \rightarrow \infty) \sim&\, \frac{\e^{-\mu}}{L^2}, \;\; \mu = \frac{2 - \tau(2)}{z} \geq 0. \label{ChalkerScaling}
\end{align}
\esub
Here $\tau(2)$ is the second multifractal exponent [Eqs.~(\ref{tauqDef}), (\ref{MFCSpec}) and (\ref{MFCDim})],
and $z$ is the dynamic critical exponent [Eq.~(\ref{DoSExpDef})--(\ref{DoSExp})]. 
The first limit in Eq.~(\ref{IPRZE}) recovers the second multifractal moment of a given
wavefunction, which defines the exponent $\tau(2)$.  
A non-zero value for this exponent implies that the state is delocalized.
The opposite limit in Eq.~(\ref{ChalkerScaling}) is (generalized)\cite{YZC-P1} 
Chalker scaling, and shows that different wavefunctions maintain mutual power-law correlations 
with respect to the energy separation. We have verified this result numerically
for the single valley Dirac fermion surface state of a class AIII TSC in Ref.~\onlinecite{YZC-P1}.   
Since multifractal wavefunctions with nearby energies exhibit a high degree of correlation, 
the effects of short-ranged interactions can be amplified (suppressed) in peaks (valleys) 
of the probability densities associated to these states.\cite{IQHP,Feigelman07,Feigelman10,Foster12-B}

The situation is completely different in an Anderson insulator. 
In the latter, two distinct wavefunctions $\psi_{\e}(\vex{r})$ and $\psi_{\e'}(\vex{r})$ close in energy 
have negligible overlap in their position space probability densities,
so that
\[
	|\psi_{\e}(\vex{r})|^2 |\psi_{\e'}(\vex{r})|^2 \sim 0 \;\; \forall \, \vex{r}.
\]
This result assumes that $|\e - \e'|$ is much smaller than the level spacing inside
a characteristic localization volume. 
As a result, weak short-ranged interactions have negligible effect in the insulating phase
(at zero temperature). See also Fig.~\ref{Fig--ChalkerWF}.

Given the above, multifractal scaling can enhance the matrix elements of interactions.\cite{IQHP,Feigelman07,Feigelman10,Foster12-B}
For a coupling strength $U$ associated to a particular four-fermion interaction, this is encoded in the ``tree level'' 
RG equation\cite{Foster12-B}
\begin{align}\label{UFlowFrame}
	\frac{d \ln U}{d l}
	=
	x_1 
	- 
	x_2^{(U)}
	+ 
	\ord{U},
\end{align}
where $x_1$ is the scaling dimension of the disorder-averaged density of states
[c.f.\ Eq.~(\ref{DoSExpDef})], while $x_2^{(U)}$ is the scaling dimension
of the four-fermion interaction operator. [These dimensions are computed
in the zero energy, (2+0)-D theory; the link to the low-energy sector of the
(2+1)-D theory is assured for (nearly) extended states by Chalker scaling.\cite{Chalker88,Chalker90,Cuevas07,YZC-P1}]
Eq.~(\ref{UFlowFrame}) is derived in Appendix \ref{App: TreeFlow}.

In the clean limit $x_2^{(U)} = 2 x_1$, independent of the details of the
interaction. Then $U$ is relevant (irrelevant) for
$x_1 < 0$ ($x_1 > 0$), corresponding to a diverging (vanishing)
density of states at zero energy [Eq.~(\ref{DoSExpDef})].
By contrast, for multifractal states in a disordered system, 
the exponent $x_2^{(U)}$ satisfies the bound $x_2^{(U)} \geq x_2$.
Here $x_2$
denotes the scaling dimension for the (disorder-averaged) second moment of the
local density of states. This is related to the second multifractal exponent
via $\tau(2) = 2(1 - x_1) + x_2$ [Eq.~(\ref{tauqDefLDoS})]. 
A four-fermion interaction $U$ that saturates the bound $x_2^{(U)} = x_2$
is \emph{maximally relevant};\cite{Cardy} whether this occurs for a particular interaction channel depends
upon the symmetry structure (in valley and spin space) of its matrix elements. 
A key difference from the clean case is that $x_2 < 2 x_1$,
because the $x_q$ characterize moments of a probability distribution.\cite{Duplantier91}
This is the mechanism by which multifractality can enhance interaction effects.

\subsection{Surface state stability and phase diagrams \label{IntStabResults}}

In subsection \ref{CIResults}, we review and elaborate upon previous results\cite{Foster12-B} for class CI.
New results for AIII and DIII appear in \ref{AIIIResults} and \ref{DIIIResults}.
These are derived in Sec.~\ref{Sec: CFT}.

\subsubsection{Spin SU(2) symmetry: Class CI \label{CIResults}}

Consider a class CI TSC with winding number $\nu = \pm 2 k$.
The non-interacting class CI surface theory is defined by the Hamiltonian in Eq.~(\ref{hci}), 
with the conformal decomposition in Eq.~(\ref{hciCFT}).
We assume that interparticle interactions preserve the bulk spin SU$(2)$ and time-reversal symmetries.
To simplify the presentation, we will also assume invariance of
the interactions under valley Sp$(2k)$ transformations, although this is not necessary. 

The structure of four-fermion interaction operators can be obtained two ways. 
A formal method is to enumerate holomorphic fermion bilinears that transform
in different irreducible representations of the spin and valley symmetry groups. 
The symmetry-allowed four fermion terms are invariant ``diagonal'' products of holomorphic and antiholomorphic
bilinears. Terms constructed in this way are automatic eigenoperators
of the conformal group, and their scaling dimensions are easily obtained. 
This is the method we employ in Sec.~\ref{Sec: CFT}.

Alternatively, one can construct interactions from products of ``physical''
operators such as currents, densities, etc. The form of the interactions can be largely inferred from 
principles of disorder-dominated quantum hydrodynamics.\cite{Finkelstein,BK,Foster06,DellAnna06,Foster08,Foster12-B}
Because spin is fully conserved, we expect a spin exchange (``triplet'') interaction of the form
$\vec{S}\cdot\vec{S}$ will play an important role, where $\vec{S}(\vex{r})$ is the spin density, 
e.g.\ $S^z = \psi^\dagger \psi / 2$ [Eq.~(\ref{AbelCurr})].
We can also have a spin current-current interaction of the form $\vec{J}_S \cdot \vec{J}_S$.
Finally, time-reversal symmetry implies that a BCS pairing interaction could induce pair formation.
As discussed in Sec.~\ref{Sec: QHEmass}, the Dirac mass operator $m(\vex{r}) = \psi^\dagger \sigh^3 \psi$
is spin SU(2) symmetric, but breaks time-reversal symmetry. A non-zero $\langle m \rangle$ opens
a gap. In a microscopic lattice model,\cite{SRL09} it can be shown\cite{Foster12-B} 
that $m(\vex{r})$ is an \emph{imaginary} spin-singlet (e.g.\ s-wave) pairing amplitude,
i.e.\ 
\[
	m(\vex{r}) \sim -i c_{\uparrow}^\dagger(\vex{r}) c_{\downarrow}^\dagger(\vex{r}) + i c_{\downarrow}(\vex{r}) c_{\uparrow}(\vex{r}),  
\]
where $c_\sigma(\vex{r})$ annihilates a lattice electron. 
We can therefore write an attractive BCS interaction as $- m^2(\vex{r})$. 
The formal and intuitive approaches to interaction operator construction 
are connected through the Fierz identities in Table~\ref{FierzTable}.

The interaction Hamiltonian is\cite{Foster12-B} 
\begin{align}\label{hcii}
	\hcii
	=&\,
	\int d^2\vex{r}
	\bigg[
	U 
	\left(
	m_a m_a - 4 \vec{S}_a \cdot \vec{S}_a
	\right)
	+
	V 	
	J^{\gamma}_{S a} \bar{J}^{\gamma}_{S a}
	\nonumber\\
	&\,
	+
	W
	\left(
	3 m_a m_a + 4 \vec{S}_a \cdot \vec{S}_a
	-
	\frac{1}{k} J^{\gamma}_{S a} \bar{J}^{\gamma}_{S a}
	\right)
	\bigg].
\end{align}
Here we have included the replica label $a$, which is summed.
In Eq.~(\ref{hcii}), $J^{\gamma}_{S a}$ ($\bar{J}^{\gamma}_{S a}$) denotes
the holomorphic (antiholomorphic) $\gamma$-spin current
($\gamma \in \{1,2,3\}$). 
The coupling constants $U,V,W$ are assigned
to particular combinations of four-fermion terms. 
These combinations are eigenoperators\cite{Foster12-B} of the conformal group
with well-defined scaling dimensions at the disordered,
non-interacting fixed point. We note that the minimal case 
$k = 1$ (two valleys) is special, in that the $W$-channel
interaction does not exist. For that case only,
$J^{\gamma}_{S a} \bar{J}^{\gamma}_{S a} = 3 m_a m_a + 4 \vec{S}_a \cdot \vec{S}_a$.

\begin{figure}[b]
   \includegraphics[width=0.35\textwidth]{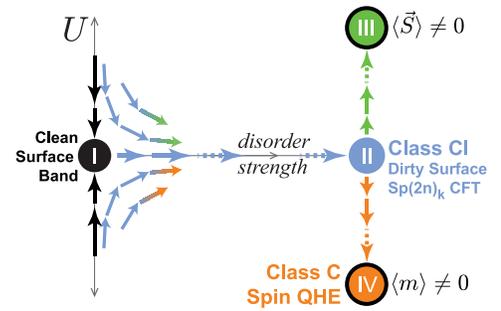}
   \caption{
	Phase portrait sketch for the surface physics 
	of a 3D class CI topological superconductor. 
	The vertical axis is the interaction 
	strength $U$ [Eq.~(\ref{hcii})], while the horizontal axis measures
	non-magnetic disorder. Although the non-interacting system has a disorder-stabilized
	phase with delocalized (``protected'') surface states {\bf (II)},
	it is destroyed by arbitrarily weak interactions [Eq.~(\ref{CITreeFlowU})].
	Instead, at zero temperature, we expect that the surface exhibits 
	broken spin symmetry [$U > 0$ $\Rightarrow$ {\bf (III)}],
	or the spin quantum Hall effect [$U < 0$ $\Rightarrow$ {\bf (IV)}].
	In either scenario, interactions break time-reversal symmetry spontaneously. 
	\label{Fig--CIPhaseDiag}
	}
\end{figure}

Using the Sp$(2 n)_k$ CFT, one can compute the scaling dimensions
$\{x_2^{(U,V,W)}\}$. 
Combining each of these with $x_1$ [Eqs.~(\ref{zDoS}) and (\ref{DoSExp})] in Eq.~(\ref{UFlowFrame}),
one finds\cite{Foster12-B}
\bsub\label{CITreeFlow}
\begin{align}
	\frac{d U}{d l}
	=&\,
	\frac{1}{2(k+1)} U
	+ 
	\ord{A^2,A B},
	\label{CITreeFlowU}
	\\
	\frac{d V}{d l} 
	=&\, 
	- 
	\frac{(4k + 3)}{2(k+1)} V
	+
	\ord{A^2,A B},
	\label{CITreeFlowV}
	\\
	\frac{d W}{d l} 
	=&\, 
	- 
	\frac{3}{2(k+1)} W	
	+
	\ord{A^2,A B},
	\label{CITreeFlowW}
\end{align}
\esub
where $A,B \in \{U,V,W\}$.
These equations imply that $U$ is relevant for \emph{any} $k$,
while $V$ and $W$ are always irrelevant.

In the clean limit, all short-ranged interactions are equally 
irrelevant, so that (e.g.)
\begin{align}\label{UFlowClean}
	\frac{d U}{d l} = - U + \ord{A^2,A B}.
\end{align}
Eq.~(\ref{CITreeFlowU}) implies that disorder promotes this interaction 
to a relevant perturbation to the non-interacting TSC surface theory.
This can be understood as a strong multifractal enhancement of the
matrix elements in the $U$-interaction channel, as discussed above in 
Sec.~\ref{IntStab}. 

To be precise, the enhancement of $U$ is due to the fact that $x_2^{(U)} = x_2$, 
i.e.\ that the dimension of the interaction operator is the same as that of the
second LDoS moment (averaged over disorder realizations). 
The latter is related to the multifractal dimension $\tau(2)$ 
[Eq.~(\ref{tauqDefLDoS})]. (See also the discussion in Sec.~\ref{IntStab}, above). 
For a dirty class CI surface, the conformally-invariant fixed point has
$x_2 = 0$ for all $k$, while $x_1 = 1/2(k+1)$.\cite{footnote--Sym}
By contrast, the other interactions have $x_2^{(V,W)} > x_2$
such that these remain irrelevant as in the clean limit. 
[In the clean limit, $x_2 = 2 x_1 = 2$, so that Eq.~(\ref{UFlowFrame})
reduces to Eq.~(\ref{UFlowClean}).]

Because $U$ couples to the difference of repulsive 
(positive-definite) spin-spin $\vec{S}\cdot\vec{S}$ and
cooper pairing $m^2$ operators, we interpret the 
RG flow away from the non-interacting CFT to imply
that arbitrarily weak interactions break time-reversal 
symmetry spontaneously.
The initial sign of $U$ depends upon microscopic details,
but a flow $U \rightarrow + \infty$ signals a
magnetic instability in which we expect $\langle \vec{S} \rangle \neq 0$,
at least in local regions.
A flow $U \rightarrow - \infty$ signals an instability
towards imaginary pairing of surface state quasiparticles,
leading to a non-zero Dirac mass $\langle m \rangle \neq 0$
and the surface spin quantum Hall effect, as discussed
above in Sec.~\ref{Sec: QHEmass}.
These results for class CI are summarized in Fig.~\ref{Fig--CIPhaseDiag}

The dimensions in Eq.~(\ref{CITreeFlow}) are exact, and were
obtained using the representation theory of the Sp$(2 n)_k$ affine Lie algebra.
A drawback of this approach is that it is very difficult to compute
higher order (``loop'') corrections to these equations in the interaction 
strengths, or to determine the backreaction of interactions upon the spin or 
thermal conductance. The problem is that these corrections entail virtual frequency 
integrations, i.e.\ the evaluation of correlation functions at one or more non-zero frequencies. 
Formally, these frequencies are relevant perturbations to the (2+0)-D
CFT. In principle, one could attempt a version of RG-improved conformal
perturbation theory to construct these objects, but such a project is likely
to run aground due to infrared divergences.

Fortunately, there is an easier way forward, at least in the limit
of large topological winding numbers $|\nu| = 2 k$. 
For $k \gg 1$, the non-abelian
bosonization of this CFT [Eq.~(\ref{SWZNW}) with $\lambda = 1/k$]
becomes a weakly-coupled non-linear sigma model, which is amenable
to a perturbative RG approach (with $1/k$ as the small parameter). 
We can generalize this to a (2+1)-D imaginary time version that
allows direct incorporation of the interactions. 
The result is a class CI Finkel'stein non-linear sigma model\cite{Finkelstein,BK,Foster06,DellAnna06}
with a Wess-Zumino-Novikov-Witten term (WZNW-FNLsM). 
The action for this theory is 
\begin{align}\label{CINLsM}
\begin{aligned}
	S_{\textrm{CI}}
	=&\,
	\frac{1}{\lambda}
	\int
	\frac{d^2\vex{r}}{8 \pi}
	\tr\left[
	\Nabla \q^\dagger 
	\cdot
	\Nabla \q 
	\right]
	+
	k 
	\WZNW
	\\
	&\,
	-
	\hf
	\int
	d^2\vex{r}
	\,
	\tr \!
	\left[
	\hat{\omega}_N
	\left(
	\q + \q^\dagger
	\right)
	\right]	
	\\
	&\,
	-
	\int
	d\tau d^2\vex{r}
	\left(
	4 \Gamma_t \,
	\vec{\mathcal{S}}_a
	\cdot
	\vec{\mathcal{S}}_a
	+
	\Gamma_c \,
	\mathcal{M}_a
	\mathcal{M}_a
	\right),
\end{aligned}
\end{align}
where $\WZNW$ is the Wess-Zumino-Novikov-Witten term [Eq.~(\ref{WZNW})].
Here 
$\hat{Q}(\vex{r}) \rightarrow Q_{a,b}^{s,s'}(\omega_n,\omega_n';\vex{r})$
is a dynamical matrix field with indices in replica ($a,b$), spin ($s,s'$),
and Matsubara frequency ($\omega_n,\omega_n'$) spaces. 
Formally $\hat{Q}$ is a group element in Sp$(2 n N)$, where $n$ ($N$) denotes
the number of replicas (Matsubara frequencies), and we are to take 
$n \rightarrow 0$, $N \rightarrow \infty$ such that $n N \rightarrow 0$. 
The first three terms in Eq.~(\ref{CINLsM}) define the non-interacting
theory. In the third term, $\hat{\omega}_N$ is a diagonal matrix of Matsubara frequencies.
This term does not appear in the (2+0)-D Wess-Zumino-Novikov-Witten CFT [Eq.~(\ref{SWZNW})], since 
the latter describes the properties at $\omega_N = 0$.

Four-fermion interactions are encoded in the final term of Eq.~(\ref{CINLsM}).
Interactions involve bilinears of $Q_{a,b}^{s,s'}(\tau,\tau')$ that are local
(diagonal) in replica and imaginary time indices.\cite{Finkelstein,BK}
The operators
\bsub\label{S,M CINLsM}
\begin{align}
	\vec{\mathcal{S}}_a(\tau,\vex{r})
	\equiv&\,	
	{\textstyle{\frac{1}{2}}}
	\tr_s \!
	\left\{
	\hat{\vec{s}}
	\left[
	\q_{a a}(\tau,\tau;\vex{r})
	+
	\q^\dagger_{a a}(\tau,\tau;\vex{r})
	\right]
	\right\},
	\\
	\mathcal{M}_a(\tau,\vex{r})
	\equiv&\,
	\tr_s \!
	\left[
	\q_{a a}(\tau,\tau;\vex{r})
	-
	\q^\dagger_{a a}(\tau,\tau;\vex{r})
	\right],
\end{align}
\esub
are the sigma model versions of the spin and BCS mass densities, respectively. 
In these equations, the trace $\tr_s$ runs over spin components; $\hat{\vec{s}}$ is the 
vector of Pauli matrices acting on spin space. 
The interaction operators conjugate to the coupling strengths $\Gamma_t$ and $\Gamma_c$
in Eq.~(\ref{CINLsM})
correspond to the spin exchange $\vec{S} \cdot \vec{S}$ and cooper
pairing $m^2$ operators in Eq.~(\ref{hcii}). 
We could also include the spin current-current
interaction, but this channel becomes strongly irrelevant in the
$k \gg 1$ limit [Eq.~(\ref{CITreeFlowV})]. 

By parameterizing $\hat{Q}$ in terms of some unconstrained coordinates, it is straightforward
to implement a perturbative RG scheme using Eq.~(\ref{CINLsM}). The non-topological version
(lacking the Wess-Zumino-Novikov-Witten term) was studied in Ref.~\onlinecite{DellAnna06}. 
At one-loop, the WZNW term modifies only the weak localization\cite{Senthil98} and Altshuler-Aronov
corrections\cite{Altshuler85} to the inverse spin conductance.\cite{WZNW-P3} The one-loop RG equations are 
given by\cite{DellAnna06,WZNW-P3}
\bsub\label{CIFlow}
\begin{align}
	\frac{d \lambda}{d l} 
	=&\,
	\lambda^2 
	\left[1 - (k \lambda)^2 \right]
	\nonumber\\
	&\, 
	\times
	\!
	\left\{
	1
	+
	3 \left[1 + \frac{1 - \gamma_t}{\gamma_t} \ln(1 - \gamma_t) \right]
	- 
	\frac{\gamma_c}{2}
	\right\},
	\label{lambdaFlowCI}
	\\
	\frac{d \gamma_t}{d l}
	=&\,
	-
	\frac{\lambda}{2}
	\gamma_c
	(1 - \gamma_t)
	(1 - 2 \gamma_t),
	\label{tFlowCI}	
	\\
	\frac{d \gamma_c}{d l}
	=&\,
	\frac{\lambda}{2}
	\left\{
	-3 \gamma_t - 2 \gamma_c
	+
	3 \gamma_c
	\left[
	\ln(1 - \gamma_t) + \gamma_t
	\right]
	\right\}
	-
	\gamma_c^2.
	\label{cFlowCI}
\end{align}
\esub
These equations are expressed in terms of the relative interaction strengths
\begin{align}
	\gamma_{t,c} \equiv \frac{4}{\pi \hf} \Gamma_{t,c},
\end{align}
where $\hf$ is the coefficient of the frequency term in Eq.~(\ref{CINLsM}).
Eq.~(\ref{CIFlow}) incorporates corrections to second homogeneous order\cite{footnote--gammac}
in $\lambda$ and $\gamma_c$, 
but includes corrections to all orders in $\gamma_t$. 

Consider first Eq.~(\ref{lambdaFlowCI}). 
In Sec.~\ref{Sec: TopProtCFT}, we argued that the parameter $\lambda$ is 
proportional to the inverse dc spin conductance. 
Then the first term (``1'') on the second line is a weak localization 
correction.\cite{Senthil98} The second and third terms are 
Altshuler-Aronov conductance corrections\cite{Altshuler85,DellAnna06}
due to the spin triplet $\gamma_t$ and Cooper channel $\gamma_c$ interactions,
respectively. 
At the conformal fixed point $\lambda = 1/k$, \emph{all} corrections 
are suppressed in Eq.~(\ref{lambdaFlowCI}).\cite{WZNW-P3} 

For non-interacting quasiparticles on the surface of a disordered class CI TSC, 
it is known\cite{Tsvelik95,Ostrovsky06} that the 
Landauer spin conductance is unmodified from its clean value 
in Eq.~(\ref{SpinLandCond}).
Eq.~(\ref{lambdaFlowCI}) implies that interaction corrections are also suppressed.
This statement is valid to lowest order in $\lambda = 1/k$.

In Ref.~\onlinecite{WZNW-P3}, we consider the lowest order Hartree and Fock 
interaction corrections to the spin conductance in the disordered 
Dirac description of the TSC surface. Without resorting to the CFT, 
we demonstrate that these vanish exactly. 
This is consistent with Eq.~(\ref{lambdaFlowCI}) only for $\lambda = 1/k$.
We therefore argue
that the Landauer spin conductance is universal and given by Eq.~(\ref{SpinLandCond}), 
so long as time-reversal symmetry is not broken (spontaneously or by external means). 
This implies that topological superconductor surface states are similar
to the chiral edge modes of the quantum Hall effect, in that the bulk 
topological winding number is encoded in a universal transport coefficient. 

In the remainder, we will set $\lambda = 1/k$ and assume that higher order
corrections do not modify this. We turn to the interaction flow equations
(\ref{tFlowCI}) and (\ref{cFlowCI}). To compare to the CFT results,
we define
\begin{align}\label{gamUW Def CI}
	\gamma_U \equiv {\textstyle{\frac{1}{4}}}(\gamma_c - 3 \gamma_t), 
	\;\; 
	\gamma_W \equiv {\textstyle{\frac{1}{4}}}(\gamma_c + \gamma_t).
\end{align}
The interaction part of Eq.~(\ref{CINLsM}) is proportional to
\begin{align}\label{CINLsMEigenInts}
	-
	\int
	d\tau d^2\vex{r}
	&
	\left[
	\gamma_U
	\left(
	\mathcal{M}_a
	\mathcal{M}_a
	- 
	4
	\vec{\mathcal{S}}_a
	\cdot
	\vec{\mathcal{S}}_a
	\right)
	\right.
	\nonumber\\	
	&\,
	\left.	
	+
	\gamma_W
	\left(
	3
	\mathcal{M}_a
	\mathcal{M}_a
	+
	4
	\vec{\mathcal{S}}_a
	\cdot
	\vec{\mathcal{S}}_a
	\right)
	\right],
\end{align}
which has the same structure as Eq.~(\ref{hcii}) with the identification 
$\gamma_{U,W} \Leftrightarrow U,W$. 
Linearizing Eqs.~(\ref{tFlowCI}) and (\ref{cFlowCI}) in the interaction strengths gives\cite{Foster12-B}
\bsub\label{CINLsMTreeFlow}
\begin{align}
	\frac{d \gamma_U}{d l}
	=&\,
	\frac{1}{2 k} \gamma_U,
	\\
	\frac{d \gamma_W}{d l}
	=&\,
	-
	\frac{3}{2 k}\gamma_W.
\end{align}
\esub
These agree with Eq.~(\ref{CITreeFlow}) to lowest order in $1/k$ for the
corresponding coupling strengths. 

\begin{figure}
   \includegraphics[width=0.43\textwidth]{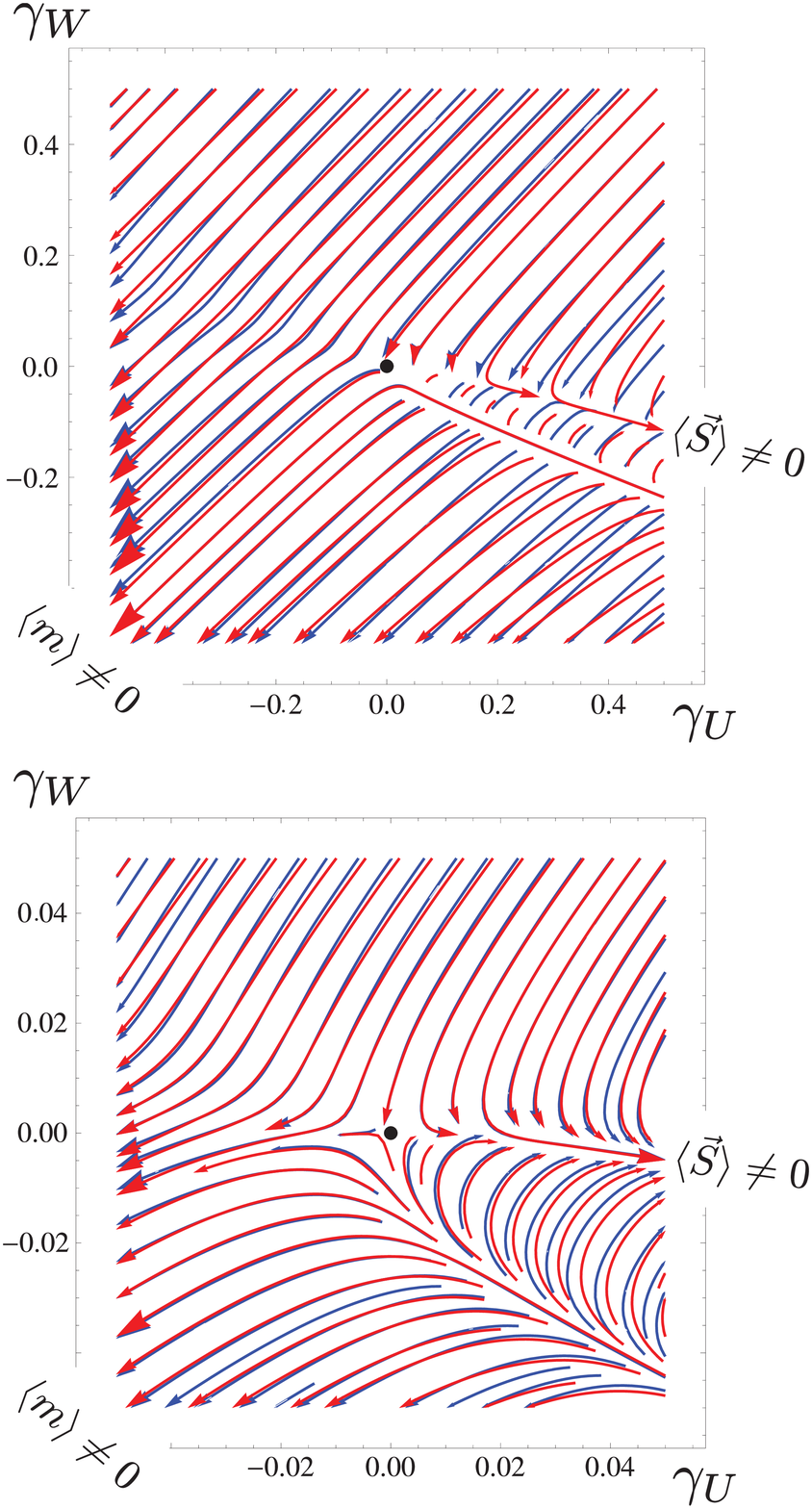}
   \caption{
	Class CI RG flows in the interaction coupling space $\gamma_U,\gamma_W$.
	The interaction $\gamma_U$ ($\gamma_W$) is a relevant (irrelevant) 
	perturbation to the non-interacting fixed point (marked by the black dot), 
	in either the CFT [Eqs.~(\ref{hcii}), (\ref{CITreeFlow})] 
	or 
	WZNW-FNLsM languages [Eqs.~(\ref{CINLsMEigenInts}), (\ref{CINLsMTreeFlow})].
	The flow in blue is the full one-loop WZNW-FNLsM result [Eqs.~(\ref{tFlowCI}) and (\ref{cFlowCI})],
	while that in red is the simplified flow in Eq.~(\ref{CINLsMFlowBCS}) that neglects all nonlinearities
	except the $-\gamma_c^2$ BCS term in Eq.~(\ref{cFlowCI}).
	In these plots we have set $k = 1/\lambda = 30$. The bottom panel
	is a zoom of the top in the vicinity of the non-interacting fixed point.
	The flows are consistent with the schematic phase diagram in 
	Fig.~\ref{Fig--CIPhaseDiag}, and further suggest that BCS pairing 
	($\gamma_U,\gamma_W \rightarrow - \infty$) is the dominant instability. 
	}
   \label{Fig--CIRG}
\end{figure}

The one-loop Eqs.~(\ref{tFlowCI}) and (\ref{cFlowCI})
possess only the non-interacting fixed point $\gamma_U = \gamma_W = 0$,
which as we have seen is unstable. 
The full non-linear structure of these equations is somewhat complicated. 
However, the main physics is captured by retaining only the
$-\gamma_c^2$ term in Eq.~(\ref{cFlowCI}), and linearizing
the rest. This term represents the BCS instability
towards imaginary Cooper pairing of the surface quasiparticles. 
Its appearance in the disordered theory is a manifestation of Anderson's theorem\cite{AndersonThm,DellAnna06}
(the interaction operator $\mathcal{M}_a \mathcal{M}_a$ is time-reversal
even, although $\mathcal{M}_a$ is odd).
In Fig.~\ref{Fig--CIRG} we plot the RG flows in the $\gamma_U$-$\gamma_W$ plane for
the full flow Eqs.~(\ref{tFlowCI}) and (\ref{cFlowCI}) (blue) and
the simplified equations (red) defined via
\bsub\label{CINLsMFlowBCS}
\begin{align}
	\frac{d \gamma_U}{d l}
	=&\,
	\frac{1}{2 k} \gamma_U
	-
	\frac{1}{4}(\gamma_U + 3 \gamma_W)^2,
	\\
	\frac{d \gamma_W}{d l}
	=&\,
	-
	\frac{3}{2 k}\gamma_W
	-
	\frac{1}{4}(\gamma_U + 3 \gamma_W)^2.
\end{align}
\esub
Eq.~(\ref{CINLsMFlowBCS}) is obtained by adding the BCS interaction
term $-\gamma_c^2$ to Eq.~(\ref{CINLsMTreeFlow}).
Fig.~\ref{Fig--CIRG} illustrates that the dominant interaction
instability is towards the direction of BCS pairing. 
In accordance with our schematic phase diagram in Fig.~\ref{Fig--CIPhaseDiag},
we expect that a flow $\gamma_c \rightarrow - \infty$  
(i.e., $\gamma_{U,W} \rightarrow - \infty$)
signals the development of the surface spin quantum Hall effect, 
with $\langle \mathcal{M}_a \rangle \sim \langle m_a \rangle \neq 0$.
(See the discussion in Sec.~\ref{Sec: QHEmass}).

\subsubsection{Spin U(1) symmetry: Class AIII \label{AIIIResults}}

For a TSC in class AIII, only a U(1) remnant of spin SU$(2)$ symmetry is preserved. 
We associate this with rotations about the $z$-axis in spin space. 
The interaction Hamiltonian is 
\begin{align}\label{haii}
	\haii
	=&\,
	\int d^2\vex{r}
	\left[
	\frac{U}{2}
	\left(
	m_a m_a 
	- 
	4 S^z_a S^z_a
	-
	\frac{4}{k}	
	J_a \bar{J}_a
	\right)
	\right.
	\nonumber\\
	&\,
	\left.
	+
	V 	
	J_a \bar{J}_a
	+
	\frac{W}{2}
	\left(
	m_a m_a + 4 S^z_a S^z_a
	\right)
	\right].
\end{align}
The structure of Eq.~(\ref{haii}) is similar to the CI case Eq.~(\ref{hcii}),
except that only the $z$-components of the spin density and current appear
($J_a \equiv J_{S a}^z$ denotes the $z$-spin current). 
Alternatively we can write 
\begin{gather}\label{haiiAlt}
	\haii
	=
	\int d^2\vex{r}
	\left[
	U_c \,
	m_a m_a 
	+
	U_t \,
	4 S^z_a S^z_a
	+
	U_j \,
	J_a \bar{J}_a
	\right],
	\nonumber\\
	U_{t,c} = {\textstyle{\frac{1}{2}}}(W \mp U),
	\;\;
	U_{j} = V - {\textstyle{\frac{2}{k}}} U.
\end{gather}

Evaluating Eq.~(\ref{UFlowFrame}) for each of the interaction
channels in Eq.~(\ref{haii}) using the U$(n)_k$ CFT, we obtain 
\bsub\label{AIIITreeFlow}
\begin{align}
	\frac{d U}{d l}
	=&\,
	\left(
	\frac{1}{k^2}
	- 	
	\lambda_A
	\right)
	U
	+ 
	\ord{A^2,A B},
	\label{AIIITreeFlowU}
	\\
	\frac{d V}{d l} 
	=&\, 
	\left(
	\frac{1 - 2k^2}{k^2} - \lambda_A
	\right)
	V
	+
	\ord{A^2,A B},
	\label{AIIITreeFlowV}
	\\
	\frac{d W}{d l} 
	=&\, 
	\left(
	-
	\frac{3 + 2 k}{k^2}
	+
	3 \lambda_A
	\right)
	W
	+
	\ord{A^2,A B},
	\label{AIIITreeFlowW}
\end{align}
\esub
where $A,B \in \{U,V,W\}$.
Eq.~(\ref{AIIITreeFlow}) is derived in Sec.~\ref{Sec: CFTAIII-Int}.

First, we note some consistency checks. Eqs.~(\ref{CITreeFlow}) with $k = 1$ and (\ref{AIIITreeFlow}) with $k = 2$
(two valleys in each case) are identical for $\lambda_A = 0$, if we remember that $W$ does not exist for class
CI when $k = 1$. Instead, $V$ in CI maps to $W$ and $V$ in AIII, which are degenerate in this
case. In the absence of the abelian spin U(1) vector potential disorder, the two valley
class CI and AIII models are the same because the valley disorder Sp(2) = SU(2).
Second, Eq.~(\ref{AIIITreeFlowV}) is consistent for $k = 1$, wherein the abelian spin U(1)
vector potential disorder $\lambda_A$ is a purely marginal perturbation to the \emph{clean} 
limit.\cite{Ludwig94} For that single valley case, only the $V$-interaction channel exists. Because this is a 
Kac-Moody current-current perturbation $x_2^{(V)} = 2$, while the scaling
dimension of the LDoS is given by $x_1 = 1 - \lambda_A$ for $k = 1$.\cite{Ludwig94}

Now we examine the criteria for surface state stability. 
The spin current-current interaction $V$ is always strongly irrelevant. 
By contrast, $U$ and $W$ may be relevant or irrelevant, depending upon the
strength $\lambda_A$ of spin current disorder. 
In particular for $\lambda_A < 1/k^2$, the $U$-interaction channel
is relevant. This is similar to CI, as the associated combination of
four-fermion terms involves a difference of repulsive spin triplet and 
Cooper pairing interactions, Eq.~(\ref{haii}).
By contrast, $W$ becomes relevant for 
$\lambda_A > (3 + 2 k)/3 k^2$. In this case the instability of the
non-interacting fixed point is mediated by $\lambda_A$, similar
to the enhancement of interactions in 2D non-topological class AIII
systems.\cite{Foster06,Foster08} 
The amplification of $W$ for sufficiently strong $\lambda_A$ is a
multifractal enhancement, but now due to the spin U(1) current
disorder. 

\begin{figure}
   \includegraphics[width=0.35\textwidth]{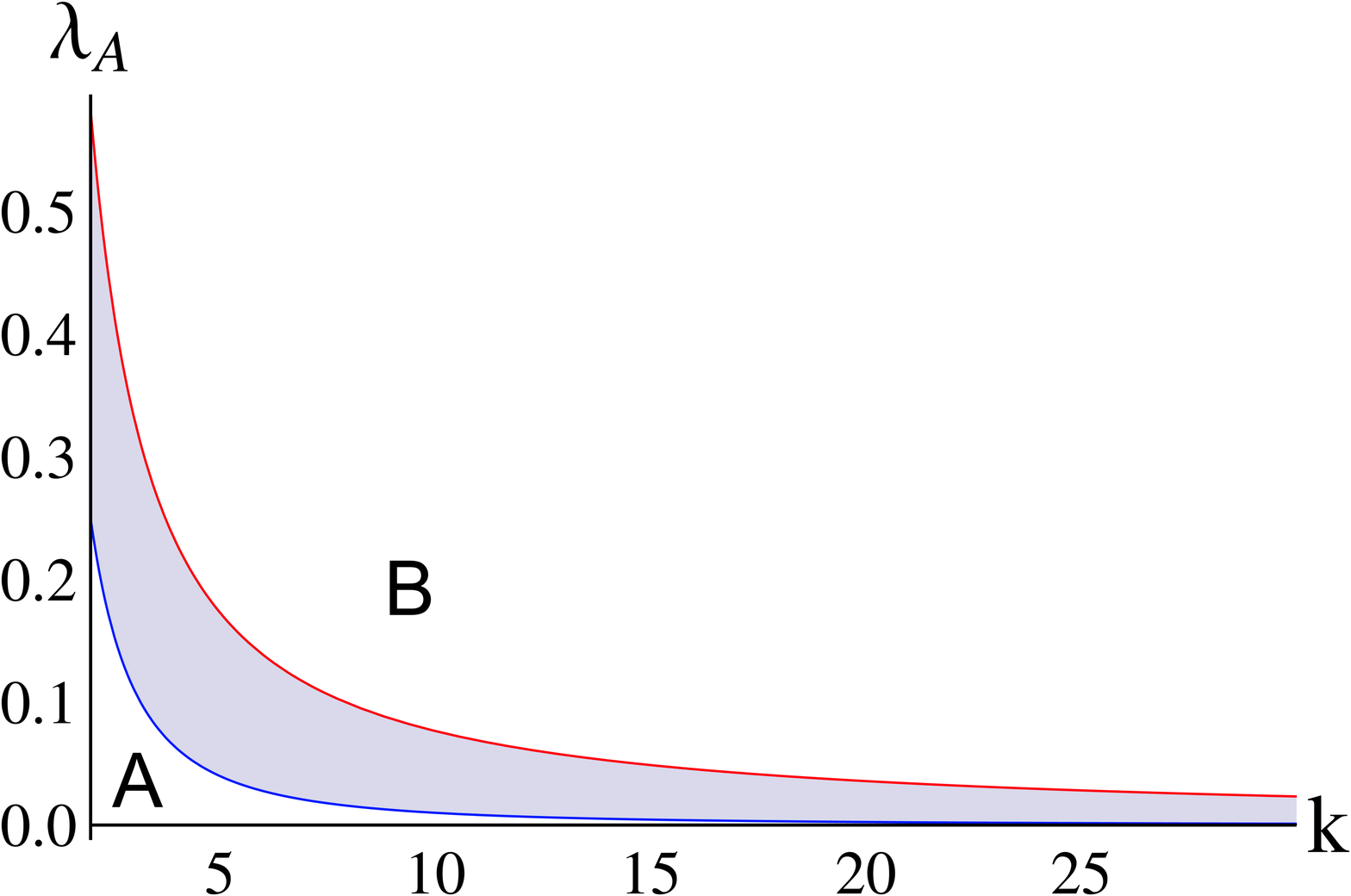}
   \caption{
	Spin current disorder-stabilized AIII TSC surface states.
	In the presence of spin U(1) current disorder $\lambda_A$,
	the disordered, non-interacting CFT description of 
	a class AIII TSC with winding number $|\nu| = k$ can be stable
	to weak interaction effects. The stable window is the
	shaded region, given by Eq.~(\ref{AIIIWindow}). 
	By contrast, the interaction channel $U$ ($W$) is 
	relevant in region {\bf A} ({\bf B})
	[Eq.~(\ref{AIIITreeFlow})].
	}
   \label{Fig--AIIIStab}
\end{figure}

We conclude that class AIII exhibits a disorder-dependent window
of stability wherein all three interactions $U,V,W$ are irrelevant,
\begin{align}\label{AIIIWindow}
	\frac{1}{k^2} < \lambda_A < \frac{3 + 2k}{3 k^2}.
\end{align}
This region is plotted in Fig.~\ref{Fig--AIIIStab}.

What happens for $\lambda_A$ outside of Eq.~(\ref{AIIIWindow})? 
Within the CFT approach, we can only say that the non-interacting
fixed point is unstable. As in class CI, in the large $k \gg 1$ limit,
the non-abelian bosonization of the U$(n)_k$ CFT becomes a weakly
coupled sigma model, with $1/k$ as a small parameter. 
Writing the (2+1)-D imaginary time version and incorporating 
interactions, we get the action
\begin{align}\label{AIIINLsM}
\begin{aligned}
	S_{\textrm{AIII}}
	=&\,
	\frac{1}{\lambda}
	\int
	\frac{d^2\vex{r}}{8 \pi}
	\tr\left[
	\Nabla \hat{U}^\dagger 
	\cdot
	\Nabla \hat{U}
	\right]
	+
	k 
	\WZNW
	\\
	&\,
	-
	\frac{\lambda_A}{\lambda^2}
	\intl{\vex{r}}
	\frac{d^2\vex{r}}{8 \pi}
	\left[
	\tr
	\left(
	\hat{U}^\dagger 
	\Nabla \hat{U}
	\right)
	\right]^2
	\\
	&\,
	-
	\hf
	\int
	d^2\vex{r}
	\,
	\tr \!
	\left[
	\hat{\omega}_N
	\left(
	\hat{U} + \hat{U}^\dagger
	\right)
	\right]	
	\\
	&\,
	-
	\sum_{a = 1}^n
	\int
	d\tau d^2\vex{r}
	\left(
	4 \Gamma_t \,
	\mathcal{S}_a^z
	\mathcal{S}_a^z
	+
	\Gamma_c \,
	\mathcal{M}_a
	\mathcal{M}_a
	\right).
\end{aligned}
\end{align}
The matrix field $\hat{U} \rightarrow U_{a,b}(\omega_n,\omega_n';\vex{r})$ is a group
element of U$(n N)$, with $n \rightarrow 0$ replicas and $N \rightarrow \infty$ Matsubara frequencies.
The interaction are bilinears of the spin and mass operators
\bsub\label{S,M AIIINLsM}
\begin{align}
	\mathcal{S}_a^z(\tau,\vex{r})
	\equiv&\,	
	{\textstyle{\frac{1}{2}}}
	\left[
	\hat{U}_{a a}(\tau,\tau;\vex{r})
	+
	\hat{U}^\dagger_{a a}(\tau,\tau;\vex{r})
	\right],
	\\
	\mathcal{M}_a(\tau,\vex{r})
	\equiv&\,
	\hat{U}_{a a}(\tau,\tau;\vex{r})
	-
	\hat{U}^\dagger_{a a}(\tau,\tau;\vex{r}).
\end{align}
\esub
As in class CI, we exclude the strongly irrelevant current-current interaction.

Relative to the non-topological case,\cite{Foster06,Foster08} the Wess-Zumino-Novikov-Witten 
term modifies only the $\lambda$ and $\lambda_A$ RG equations.
The one-loop results for the WZNW-FNLsM are\cite{Foster06,Foster08,WZNW-P3}
\bsub\label{AIIIFlow}
\begin{align}
	\frac{d \lambda}{d l} 
	=&\, 
	\lambda^2 
	\left[1 - (k \lambda)^2\right]
	\mathcal{I},
	\label{lambdaFlowAIII}
	\\
	\frac{d \lambda_A}{d l}
	=&\, 
	\lambda^2
	\left[1 - (k \lambda)^2\right]
	\left(
	1
	+
	\frac{2 \lambda_A \mathcal{I}}{\lambda}
	\right),
	\label{lambdaAFlowAIII}
	\\
	\frac{d \gamma_t}{d l}
	=&\,
	\lambda_{A} (1-\gamma_t)  \left(\gamma_t + 2 \gamma_{c} - 2 \gamma_t \gamma_{c} \right) 
	\nonumber\\
	&\,
	- 
	\lambda (1-\gamma_t) \left( \gamma_t + \gamma_{c} - 2 \gamma_t \gamma_{c} \right),
	\label{tFlowAIII}	
	\\
	\frac{d \gamma_c}{d l}
	=&\,
	\lambda_{A} \left(2 \gamma_t + \gamma_{c}\right) - \lambda \left(\gamma_t + \gamma_{c}\right)
	\nonumber\\
	&\,+ 
	\lambda \gamma_c \left[2 \ln(1 - \gamma_t) + \gamma_t \right] - 2 \gamma_{c}^{2},
	\label{cFlowAIII}
\end{align}
\esub
where
\begin{align}
	\mathcal{I} \equiv 2 \left[1 + \frac{1 - \gamma_t}{\gamma_t} \ln(1 - \gamma_t)\right] - \gamma_c
\end{align}
is the Altshuler-Aronov\cite{Altshuler85} spin conductance correction.
The relative interactions are 
\begin{align}
	\gamma_{t,c} \equiv \frac{4}{\pi \hf} \Gamma_{t,c}.
\end{align}

The non-interacting CFT has $\lambda = 1/k$. Eqs.~(\ref{lambdaFlowAIII}) and (\ref{lambdaAFlowAIII})
imply that this is a fixed point of $\lambda$ even in the presence of interactions,
and that the abelian disorder parameter $\lambda_A$ remains purely marginal in this case. 
One can say that the topology neutralizes the Altshuler-Aronov correction,
and this conclusion holds for CI [Eq.~(\ref{lambdaFlowCI})] and DIII as well [Eq.~(\ref{lambdaFlowDIII})].\cite{WZNW-P3}

In Sec.~\ref{Sec: TopProtCFT}, we noted that tuning $\lambda < 1/k$ induces runaway
flow in $\lambda_A$, in the absence of interactions [Eq.~(\ref{GLLlambdaAFlow})].\cite{GLL}
This runaway flow is known as ``Gade'' scaling,\cite{Gade,Loc} and is characteristic
of non-topological class AIII models in 2D. It is known that Gade scaling induces
a density of states divergence at zero energy,\cite{Gade,GLL,Loc} as well as strong multifractality in 
the low-energy wavefunctions. 
The runaway $\lambda_A$ flow amplifies
interactions, so that these are always relevant.\cite{Foster06,Foster08}

In Sec.~\ref{Sec: TopProtCFT}, we presented evidence that the value of $\lambda$ is pinned
to $1/k$. This includes the universality of the Landauer spin conductance in the 
absence of interactions, which is proportional to $k$. In the disordered Dirac language
of Eq.~(\ref{hai}), the lowest order Hartree and Fock interaction corrections to the conductance vanish exactly.\cite{WZNW-P3}
In what follows, we set $\lambda = 1/k$ and assume that higher order corrections do
not destabilize this. As a result, $\lambda_A$ remains a tunable, strictly marginal
parameter. 

To compare to the CFT, we linearize Eqs.~(\ref{tFlowAIII}) and (\ref{cFlowAIII}) and
express the results in terms of 
\begin{align}
	\gamma_{U,W} \equiv \gamma_c \mp \gamma_t.
\end{align}
One obtains
\bsub\label{AIIINLsMTreeFlow}
\begin{align}
	\frac{d \gamma_U}{d l}
	=&\,
	- \lambda_A \gamma_U,
	\\
	\frac{d \gamma_W}{d l}
	=&\,
	\left( - \frac{2}{k} + 3 \lambda_A \right) \gamma_W.
\end{align}
\esub
These are consistent with Eq.~(\ref{AIIITreeFlow}) to lowest order in $1/k$,
using the correspondence $\gamma_{U,W} \Leftrightarrow U,W$.

The full one-loop flow Eqs.~(\ref{tFlowAIII}) and (\ref{cFlowAIII}) possess
a non-trivial fixed point that is locally unstable (stable) for 
$\lambda_A <  \frac{2}{3 k}$ ($\lambda_A >  \frac{2}{3 k}$). 
In the latter case, this constitutes a new, interaction-stabilized
fixed point. The fixed point coupling strengths $\{\gamma_t^*,\gamma_c^*\}$
are complicated functions of $k$ and $\lambda_A$---see Eq.~(\ref{AIIIFPFull}) 
in Appendix~\ref{App: AIIIFP}. 
The key point is that the interaction-stabilized fixed point nucleates 
from the non-interacting CFT ($\{\gamma_t,\gamma_c\} = \{0,0\}$) at 
$\lambda_A = 2/3k$, and is thus fully-controlled for some range of 
$\lambda_A$. As $\lambda_A$ is increased above $2/3k$, this 
new fixed point quickly moves to large $\gamma_W$, beyond the regime 
of reliability for the weak-coupling RG.

\begin{figure}
   \includegraphics[width=0.4\textwidth]{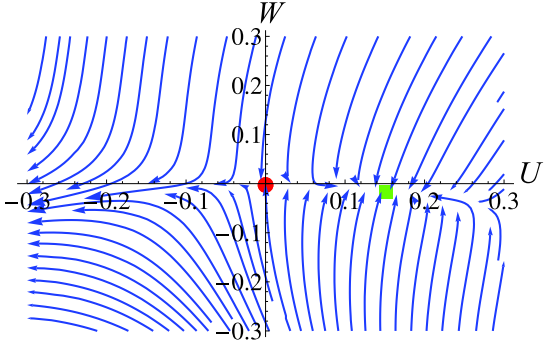}
   \caption{
	Simplified RG flow in the $U$-$W$ interaction plane for class AIII.
	Flows are plotted for Eq.~(\ref{AIIINLsMFlowBCS}).
	In this figure, $\lambda_A = 0$, and we have set $k = 4$.
	The red (green) dot denotes the non-interacting (interaction-stabilized)
	fixed point, which is unstable (stable) for 
	$\lambda_A < 1/k^2$.
	}
   \label{Fig--AIIIPhaseDiag-1}
\end{figure}

\begin{figure}
   \includegraphics[width=0.4\textwidth]{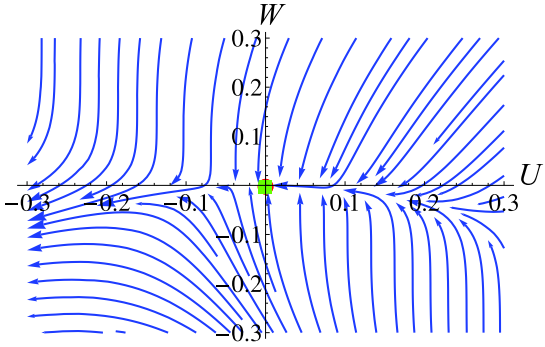}
   \caption{
	The same as Fig.~\ref{Fig--AIIIPhaseDiag-1}, but for 
	$\lambda_A = 1/k^2$.
	}
   \label{Fig--AIIIPhaseDiag-2}
\end{figure}

We note that Eq.~(\ref{AIIINLsMTreeFlow}) seems to imply that $\gamma_U$ is
irrelevant for any $\lambda_A > 0$. The CFT result in Eq.~(\ref{AIIITreeFlow})
shows that this is not correct, leading instead to the stability window in 
Eq.~(\ref{AIIIWindow}). [The discrepancy is the $1/k^2$ term in Eq.~(\ref{AIIITreeFlowU}),
which would appear in the WZNW-FNLsM treatment only at two loops.]
As in class CI, we can get a more intuitive understanding of the 
interaction plane RG flow by adding only the pure BCS term 
$-2 \gamma_c^2 = - (1/2)(\gamma_U + \gamma_W)^2 \Leftrightarrow -(1/2)(U + W)^2$
to the CFT results in Eq.~(\ref{AIIITreeFlow}), leading to
\bsub\label{AIIINLsMFlowBCS}
\begin{align}
	\frac{d U}{d l}
	=&\,
	\left(
	\frac{1}{k^2}
	- 	
	\lambda_A
	\right)
	U
	-
	\frac{1}{2}
	(U + W)^2,	
	\\
	\frac{d W}{d l} 
	=&\, 
	\left(
	-
	\frac{3 + 2 k}{k^2}
	+
	3 \lambda_A
	\right)
	W
	-
	\frac{1}{2}
	(U + W)^2.
\end{align}
\esub

\begin{figure}
   \includegraphics[width=0.4\textwidth]{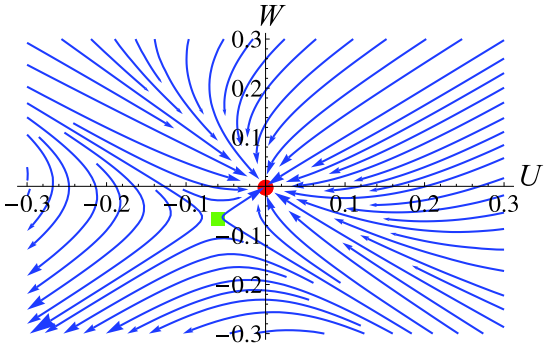}
   \caption{
	The same as Fig.~\ref{Fig--AIIIPhaseDiag-1}, but for 
	$1/k^2 < \lambda_A < (3+2k)/3k^2$.	
	In this case the non-interacting fixed point (red) 
	is stable, while the non-trivial fixed point (green)
	is unstable}
   \label{Fig--AIIIPhaseDiag-3}
\end{figure}

\begin{figure}
   \includegraphics[width=0.4\textwidth]{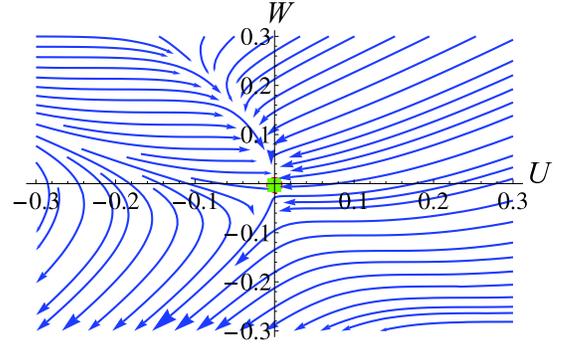}
   \caption{
	The same as Fig.~\ref{Fig--AIIIPhaseDiag-1}, but for 
	$\lambda_A = (3+2k)/3k^2$.
	}
   \label{Fig--AIIIPhaseDiag-4}
\end{figure}

\begin{figure}
   \includegraphics[width=0.4\textwidth]{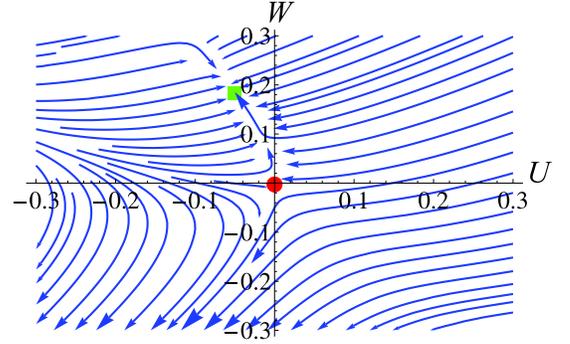}
   \caption{
	The same as Fig.~\ref{Fig--AIIIPhaseDiag-1}, but for 
	$\lambda_A > (3+2k)/3k^2$.	
	In this case the non-interacting fixed point (red) 
	is unstable, while the non-trivial fixed point (green)
	is stable.
	}
	\label{Fig--AIIIPhaseDiag-5}
\end{figure}

Eq.~(\ref{AIIINLsMFlowBCS}) neglects most of the non-linear interaction terms
in Eqs.~(\ref{tFlowAIII}) and (\ref{cFlowAIII}), but  
gives the same qualitative behavior as these for $\lambda_A > 1/k^2$.
Eq.~(\ref{AIIINLsMFlowBCS}) possesses a non-trivial fixed point
for generic $\lambda_A < (k+1)/k^2$. The non-trivial fixed point is unstable
for $\lambda_A$ within the range in Eq.~(\ref{AIIIWindow}), wherein
the non-interacting CFT ($U = W = 0$) is stable. At the boundaries of this range,
the non-trivial fixed point merges with the non-interacting theory.
Like the full Eqs.~(\ref{tFlowAIII}) and (\ref{cFlowAIII}), the non-trivial
fixed point is \emph{stable} for $\lambda_A > (3+2k)/3k^2$, but quickly
moves to strong coupling as $\lambda_A$ is increased beyond this threshold value.
A new feature of Eq.~(\ref{AIIINLsMFlowBCS}) is that the non-trivial fixed
point is also stable for $\lambda_A < 1/k^2$. 

\begin{figure}
   \includegraphics[width=0.4\textwidth]{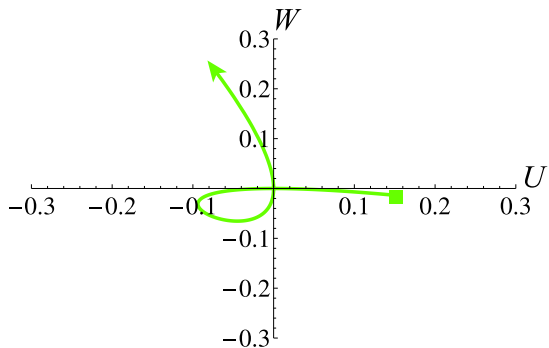}
   \caption{Interaction plane trajectory of the non-trivial fixed point
	shown in Figs.~\ref{Fig--AIIIPhaseDiag-1}--\ref{Fig--AIIIPhaseDiag-5} as a function
	of increasing $\lambda_A$. 
	}
   \label{Fig--AIIIFlowTraj}
\end{figure}

In Figs.~\ref{Fig--AIIIPhaseDiag-1}--\ref{Fig--AIIIPhaseDiag-5}, we plot
the flow fields in the $U$-$W$ interaction plane corresponding to 
Eq.~(\ref{AIIINLsMFlowBCS}), for different values of the abelian
disorder strength $\lambda_A$. The main takeaway is that 
for $\lambda_A < (1+k)/k^2$
there is 
always a critically delocalized, time-reversal invariant fixed point.
For $\lambda_A$ in the
range given by Eq.~(\ref{AIIIWindow}), this is the non-interacting
U$(n)_k$ CFT. Outside of this range, one finds new interaction-stabilized
fixed points. The latter merge with the non-interacting one at
the boundaries of Eq.~(\ref{AIIIWindow}), Figs.~\ref{Fig--AIIIPhaseDiag-2} and \ref{Fig--AIIIPhaseDiag-4}. 
The trajectory of the non-trivial fixed point as a function of
increasing $\lambda_A$ is shown in Fig.~\ref{Fig--AIIIFlowTraj}.
The flow fields in the $\{\gamma_U,\gamma_V\}$ plane for the 
full one-loop results in Eqs.~(\ref{tFlowAIII}) and (\ref{cFlowAIII})
are qualitatively the same as those shown in these figures,
except that the non-interacting fixed point remains stable
for arbitrarily small $\lambda_A$ and the non-trivial
fixed point merges with this at $\lambda_A = 0$.

Comparing the interaction flows for class AIII in Figs.~\ref{Fig--AIIIPhaseDiag-1}--\ref{Fig--AIIIPhaseDiag-5}
to class CI in Fig.~\ref{Fig--CIRG}, it is apparent that the magnetic instability in the latter
($\langle \vec{S} \rangle \neq 0$) has been replaced by a critically delocalized fixed point.
This is either the non-interacting conformal fixed point for $\lambda_A$ in the window bounded by
Eq.~(\ref{AIIIWindow}), or an interaction-stabilized one outside of this. On the other hand, the instability to surface
state Cooper pairing ($U,W \rightarrow - \infty$) is common to both AIII and CI. 
We conclude that surface states of TSCs in classes AIII and CI are vulnerable to 
the formation of gapped surface quantum Hall phases with spontaneously broken time-reversal symmetry.
As discussed in Sec.~\ref{IntStab}, the mechanism is the multifractal enhancement of 
interaction matrix elements, due to the disorder. Unlike class CI, however,
class AIII possesses a range of disorder and interaction strengths wherein time-reversal
symmetry and critical delocalization are preserved.


\subsubsection{No spin symmetry: Class DIII \label{DIIIResults}}

Finally we turn to class DIII. Bulk topological superconductors in this 
class are characterized by the integer-valued winding number $\nu$.
As discussed in the Introduction, the minimal case $k \equiv |\nu| = 1$
corresponding to a single Majorana fermion band is manifestly robust to
both disorder and interactions, in the sense that both are strongly irrelevant.
The two valley $k = 2$ version is equivalent to class AIII with $k = 1$, 
because this theory exhibits an emergent valley U(1) symmetry. 

For $k \geq 3$, class DIII admits non-abelian valley vector potential
disorder. 
Critically delocalized wavefunctions are described by the
SO$(n)_k$ CFT [Eq.~(\ref{Embed})], with multifractal 
local density of states (LDoS)
fluctuations captured by Eq.~(\ref{MFCSpec}) and (\ref{MFCDim}). 

Since spin is not conserved in class DIII, the only valley-invariant
short-ranged interaction is the BCS pairing term
[c.f.\ Eqs.~(\ref{hcii}) and (\ref{haii})].
The interaction Hamiltonian is
\begin{align}\label{hdii}
	\hdii
	=&\,
	U
	\int d^2\vex{r} 
	\,
	m_a m_a, 
\end{align}
where $m_a = \chi^\T \hat{M}_p \, \sigh^3 \chi$ is the Majorana mass
operator [Eq.~(\ref{mass})].
Using the SO$(n)_k$ CFT, we compute $x_2^{(U)}$ and evaluate
Eq.~(\ref{UFlowFrame}). The result is
\begin{align}
	\frac{d U}{d l}
	=&\,
	-
	\frac{1}{(k - 2)} U
	+ 
	\ord{U^2}.
	\label{DIIITreeFlowU}
\end{align}
Eq.~(\ref{DIIITreeFlowU}) is derived in Sec.~\ref{Sec: CFTDIII-Int}.

Unlike classes CI and AIII, we conclude that the BCS pairing channel
is \emph{always irrelevant} in class DIII. For sufficiently weak interactions,
the surface physics of a disordered DIII TSC with $k \geq 3$ is governed
by the SO$(n)_k$ CFT, which gives universal predictions for LDoS statistics and
which should exhibit the universal thermal conductance
given by Eq.~(\ref{ThermalCond}).

Eq.~(\ref{DIIITreeFlowU}) arises despite the strong disorder enhancement of the
average density of states (DoS) $\nu(\e)$. Eqs.~(\ref{DoSExpDef}) and (\ref{DoSExp}) imply that 
$\nu(\e) \sim |\e|^{-1/(2k - 3)}$, which diverges as $\e \rightarrow 0$. 
By contrast, the average DoS in class CI always vanishes, while it may vanish or
diverge in AIII. We would naively expect interactions to play the strongest
role in class DIII. This does not occur, however, because the interaction operator
$m_a m_a$ in Eq.~(\ref{hdii}) does not have the same symmetry structure as the 
second LDoS moment. 
In class DIII, the $q$th LDoS moment has the multifractal scaling dimension
\begin{align}\label{DeltaqDIII}
	x_q = - \frac{q^2}{(k - 2)},
\end{align}
so that $x_2 = -4/(k-2) < 2 x_1 = -2/(k-2)$. 
[Eq.~(\ref{DeltaqDIII}) is derived in Sec.~\ref{Sec: CFTDIII-MFC}].
In evaluating Eq.~(\ref{UFlowFrame}), we must determine $x_2^{(U)}$,
the scaling dimension of the four fermion interaction. 
In class CI this is equal to $x_2$, implying that
the $U$-channel interaction [as defined by Eq.~(\ref{hcii})] 
is maximally relevant. The story is different in class DIII, and
$x_2^{(U)} = 0 > x_2$, as shown in Sec.~\ref{Sec: CFTDIII-Int}.

Physically, this means that while the average density of states is amplified,
wavefunction multifractality does not enhance the matrix elements of the interactions.
This is a key concept: it is not enough to say that one has critically delocalized
wavefunctions with multifractal LDoS fluctuations, strongly correlated in energy
by Chalker scaling. (See the discussion in Sec.~\ref{IntStab}.) 
Rather, one also requires that the particular interaction channel of interest
has a non-vanishing overlap with the second LDoS moment, in the sense of 
common symmetry structures (in spin, valley, and replica indices). 
In class DIII the interaction operator has an orthogonal structure to the 
multifractal moment: the former is symmetric in replica indices, while
the latter can be taken to be antisymmetric (Sec.~\ref{Sec: CFTDIII-MFC}).
A similar disparity between interactions and multifractal moments arises
for a 2D unitary metal with short-ranged interactions, which are also
irrelevant.\cite{BK}\\

One can also formulate a WZNW-FNLsM for class DIII.\cite{WZNW-P3} Because the interaction
$U$ is irrelevant, the only new information this provides concerns the
beta function for the
ratio of the temperature to the thermal conductance,\cite{WZNW-P3}
proportional to the parameter $\lambda$ which satisfies the one-loop RG equation
\begin{align}\label{lambdaFlowDIII}
	\frac{d \lambda}{d l}
	=&\,
	\lambda^2
	[1 - (k \lambda)^2]
	\left(
	-
	2
	-
	\gamma_c
	\right).
\end{align}
As with classes CI and AIII, both the weak antilocalization\cite{Senthil00} (``$-2$'') 
and Altshuler-Aronov (``$-\gamma_c$'')
corrections to the conductance are suppressed at the conformal fixed point $\lambda = 1/k$.\cite{WZNW-P3}
Compare to Eqs.~(\ref{lambdaFlowCI}) (CI), (\ref{lambdaFlowAIII}), and (\ref{lambdaAFlowAIII}) (AIII).


\section{Conformal analysis \label{Sec: CFT}}

In this section we derive the conformal field theory results discussed in Sec.~\ref{Sec: Results}.
The multifractal spectrum of LDoS fluctuations is computed in Secs.~\ref{Sec: CFTAIII-MFC} and \ref{Sec: CFTDIII-MFC} for classes
AIII and DIII.
We enumerate interparticle interaction operators for these classes and compute their scaling dimensions in 
Secs.~\ref{Sec: CFTAIII-Int} and \ref{Sec: CFTDIII-Int}.
The analysis for class CI has appeared elsewhere.\cite{Foster12-B}

\subsection{Class AIII}

\subsubsection{Density of states and multifractal spectrum \label{Sec: CFTAIII-MFC}}

The low-energy effective CFT for a class AIII TSC surface state with $k$ valleys
is the U$(1) \, \oplus \, $SU$(n)_k$ affine Lie algebra [Eq.~(\ref{Embed})]. 
Here $n \rightarrow 0$ is the number of replicas. 
The (2+0)-D action may be written as [C.f.\ Eq.~(\ref{SWZNW})]
\begin{align}\label{AIIIWZNW}
\begin{aligned}
	S_{\mathrm{AIII}}^{\puprm{WZNW}}
	=&\,
	n k \left(1 + n k \lambda_A\right)
	\int
	\frac{d^2\vex{r}}{8 \pi}
	\Nabla \phi \cdot \Nabla \phi
	\\
	&\,
	+
	k
	\int
	\frac{d^2\vex{r}}{8 \pi}
	\tr\left[
	\Nabla \hat{Q}^{\dagger}
	\cdot
	\Nabla \hat{Q}
	\right]
	+
	k \WZNW,
\end{aligned}
\end{align}
where $\hat{Q}$ is an SU$(n)$ group element acting on the fundamental representation
of the replica space, $\WZNW$ is the Wess-Zumino-Novikov-Witten term [Eq.~(\ref{WZNW})],
and $\phi$ denotes the free boson field [abelian bosonization of the spin U(1) sector].

The appearance of the abelian disorder parameter $\lambda_A$ in Eq.~(\ref{AIIIWZNW}) 
can be understood as follows. Even for $\lambda_A = 0$, the free boson $\phi$ still plays
an important role, in that holomorphic primary fields of the parent U$(n k)_1$ theory are products
of vertex operators $:\exp(\pm i \beta \varphi):$ and primary fields from the SU$(n)_k$ and SU$(k)_n$ 
replica and valley constituents.\cite{Nersesyan94} Here $\varphi$ denotes the holomorphic component of $\phi$,
\[
	\phi(\vex{r}) = \varphi(z) + \bar{\varphi}(\bar{z}).
\]
For example, the holomorphic free fermion field $L$ in Eq.~(\ref{haioDef}) can be understood
(roughly)\cite{QCD} as the product
\begin{align}
	L_{v,a}(z) = : \exp\left[i \sqrt{4 \pi} \varphi(z)\right] : \, \Omega^{\puprm{R}}_{1,a}(z) \, \Omega^{\puprm{V}}_{1,v}(z),
\end{align}
where $\Omega^{\puprm{R}}_{1}(z)$ and  $\Omega^{\puprm{V}}_{1}(z)$
denote the primary fields in the fundamental representation of the 
replica SU$(n)_k$ and valley SU$(k)_n$ sectors, respectively. 
In the clean theory [Eq.~(\ref{hai}) with $\vex{A}_i = \vex{A} = 0$],  
$L(z)$ has the holomorphic dimension\cite{CFT}
\begin{align}\label{AIII-h1Def}
	h_1 \equiv&\,
	h_1^{\pupsq{\textrm{U}(1)}} + h_1^{\pupsq{\textrm{SU}(n)_k}} + h_1^{\pupsq{\textrm{SU}(k)_n}}
	\nonumber\\
	=&\,
	\frac{1}{2 n k} + \frac{n^2 - 1 }{2 n(n+k)} + \frac{k^2 - 1 }{2 k(n+k)}
	= \frac{1}{2}.
\end{align}
Here we have used the normalization of the $\phi$ action in Eq.~(\ref{AIIIWZNW}) with $\lambda_A = 0$. 

The abelian vector potential $\vex{A}$ in Eq.~(\ref{hai}) couples to the spin U(1)
current $\vex{J}$. With our normalization, this bosonizes to $(n k/\sqrt{4 \pi}) i \Nabla \phi$. 
Averaging over the disorder gives the $\lambda_A$ term in Eq.~(\ref{AIIIWZNW}). 

In the presence of disorder, the SU$(k)_n$ sector ``localizes.'' 
As a result, the dimension of $L(z)$ in the low-energy U$(n)_k$ CFT is reduced to
\begin{align}
	h_1 = \frac{1}{2 n k(1 + \lambda_A n k)} + \frac{n^2 - 1 }{2 n(n+k)}.
\end{align}
The LDoS operator is $R^\dagger_a L_a + L^\dagger_a R_a$, which has 
dimension 
\begin{align}\label{LDoSAIII-CFT}
	x_1 = 2 h_1 = \frac{1}{k^2} - \lambda_A,
\end{align}
where we have taken the replica $n \rightarrow 0$ limit. 
This result was originally obtained in Ref.~\onlinecite{Nersesyan94}.

Higher LDoS moments are not conformal group eigenoperators. However,
the $q$th LDoS moment has a most relevant component that corresponds
to the primary field with weight $\omega_q$, which is the completely antisymmetric
tensor representation with $q$ indices. 
Here the symbol $\omega_p$, $p \in \{1,2,\ldots,r\}$ denotes the $p$th fundamental weight
for a rank $r$ Lie algebra.\cite{CFT}
As a result, the $q$th moment has the scaling dimension
\begin{align}\label{DeltaqAIII}
	x_q
	=&\,
	2
	\left(h_q^{\pupsq{\textrm{U}(1)}} + h_q^{\pupsq{\textrm{SU}(n)_k}}\right)
	\nonumber\\
	=&\,
	q 
	- 
	\left(\frac{k - 1}{k^2}\right)(k q+q^2) - \lambda_A q^2,
\end{align}
where we have taken the replica limit. 
In Eq.~(\ref{DeltaqAIII}), for non-zero $n$ one has
\[
	h_q^{\pupsq{\textrm{U}(1)}}
	=
	\frac{q^2}{2nk(1 + \lambda_A n k)},
	\;\;	
	h_q^{\pupsq{\textrm{SU}(n)_k}}
	=
	\frac{q(n-q)(n+1)}{2n(n+k)}.
\]
Eq.~(\ref{DeltaqAIII}) was originally obtained in Refs.~\onlinecite{Mudry96,Caux96}.

The class AIII multifractal spectrum in Eq.~(\ref{MFCSpec}) and (\ref{MFCDim}) obtains from 
combining Eqs.~(\ref{tauqDefLDoS}) and (\ref{DeltaqAIII}).

\subsubsection{Interaction operators \label{Sec: CFTAIII-Int}}

We work in the language of the holomorphic decomposition in 
Eq.~(\ref{haiCFT}). A generic rotationally invariant, spin U$(1)$ 
invariant interaction can be expressed as a linear combination of 
\bsub
\begin{align}
	\mathcal{O}_{a, \, v_1 v_3}^{(\msf{A})\,v_2 v_4}
	\equiv& \,
	R_{v_1,a}(\bar{z})\, 
	R^{\dagger \, v_2, a}(\bar{z})\;
	L_{v_3,a}(z)\, 
	L^{\dagger \, v_4, a}(z),
	\\
	\mathcal{O}_{a, \, v_3 v_4}^{(\msf{B})\,v_1 v_2}
	\equiv& \,
	R^{\dagger \, v_1,a}(\bar{z})\, 
	R^{\dagger \, v_2, a}(\bar{z})\; 
	L_{v_3,a}(z)\, 
	L_{v_4, a}(z),
	\\
	\mathcal{O}_{a, \, v_1 v_2}^{(\msf{C})\,v_3 v_4}
	\equiv& \,
	R_{v_1,a}(\bar{z})\, 
	R_{v_2, a}(\bar{z})\;
	L^{\dagger \, v_3, a}(z)\, 
	L^{\dagger \, v_4, a}(z).
\end{align}
\esub
Here and in what follows, no sum on the replica index $a$ is implied.

{\flushleft{\underline{Interaction type $\msf{A}$}}}\\

We consider first $\mathcal{O}_{a}^{(\msf{A})}$, with holomorphic component 
$L_{v,a}(z)\, L^{\dagger \, v', a}(z) $. 
The traceless, replica-resolved tensor
\[
	L_{v,a}\, L^{\dagger \, v',a} 
	- 
	\frac{1}{k} \delta_{v}^{v'} \sum_{v''} L_{v'',a}\, L^{\dagger \, v'',a}
\]
transforms in the adjoint representation of the valley SU$(k)$; because it is replica-resolved, this
bilinear cannot be a valley Kac-Moody current. For SU$(N)_q$ with $q \geq 2$, the adjoint representation can
be associated to a primary field with
the affine weight $(q-2)\hat{\omega}_0 + \hat{\omega}_1 + \hat{\omega}_{N-1}$, with holomorphic dimension
\begin{align}
	h_{(1 0 \cdots 0 1)} 
	=& 
	\frac{N}{N + q}.
\end{align}
[The indices in the $(\cdots)$ subscript denote (non-affine) Dynkin coefficients.]
The product of SU$(n)_k$ and SU$(k)_n$ adjoint representation primary fields has the holomorphic dimension
\[
	\frac{n}{n + k} + \frac{k}{n + k} = 1,
\]
as expected. In the conformal remnant theory U$(1) \, \oplus \, $SU$(n)_k$ [Eq.~(\ref{Embed})],
the corresponding interaction operator (diagonal primary field $\equiv \mathcal{O}_{a}^{(\msf{A},1)}$) 
carries scaling dimension $x_2^{(\msf{A},1)} = 0$ in the replica $n \rightarrow 0$ limit. 
We can write this interaction operator as 
\begin{align}\label{OA1}
	\mathcal{O}_{a}^{(\msf{A},1)} 
	\equiv&\, 
	2 J_{\kappa a}^i \bar{J}_{\kappa a}^i,
	\quad	
	x_2^{(\msf{A},1)} = 0,
\end{align}
where 
\begin{align}
	J_{\kappa a}^i \equiv L^{\dagger \, a} \hat{t}_\kappa^i L_a,\;\;
	\bar{J}_{\kappa a}^i \equiv R^{\dagger \, a} \hat{t}_\kappa^i R_a
\end{align}
denote replica-resolved valley SU$(k)$ currents. 

We also define the valley singlet operators 
\begin{align}
	J_a \equiv \sum_{v}  L^{\dagger \, v, a} L_{v,a},\;\;
	\bar{J}_a \equiv \sum_{v} R^{\dagger \, v, a} R_{v,a}.
\end{align}
These are replica Kac-Moody currents. The corresponding 
interaction operator has dimension $2$. We denote this as 
\begin{align}\label{OA2}
	\mathcal{O}_{a}^{(\msf{A},2)} 
	\equiv&\, 
	J_{a} \bar{J}_{a},
	\quad
	x_2^{(\msf{A},2)} = 2.
\end{align}

{\flushleft{\underline{Interaction types $\msf{B,C}$}}}\\

We consider next $\mathcal{O}_{a}^{(\msf{B})}$.
The relevant holomorphic bilinear is $ L_{v_3,a}(z) L_{v_4, a}(z) $.
This must be antisymmetrized in valley indices (due to Fermi statistics),
while the replica indices belong to the symmetric representation.  
The relevant affine weight for the antisymmetric (symmetric) representation is 
$(q - 1)\hat{\omega}_0 + \hat{\omega}_2$ [$(q - 2)\hat{\omega}_0 + 2\hat{\omega}_1$]. 
In SU$(N)_q$,
\begin{align}
\begin{aligned}
	h_{(0 1 0 \cdots 0)} 
	=& 
	\frac{(N - 2)(N + 1)}{N (N + q)},
	\\
	h_{(2 0 0 \cdots 0)} 
	=& 
	\frac{(N + 2)(N-1)}{N(N + q)}.
\end{aligned}
\end{align}

For $\lambda_A = 0$, the product of U$(1)$, $(2 0 0 \cdots 0)$ SU$(n)_k$,
and $(0 1 0 \cdots 0)$ SU$(k)_n$ primary fields must carry conformal
dimension $h = 1$. This determines the $U(1)$ dimension $h^{\pupsq{\textrm{U}(1)}}$ via
\begin{align}
	\left.h^{\pupsq{\textrm{U}(1)}}\right|_{\lambda_A = 0} 
	=&  
	1 
	- 
	h_{(2 0 0 \cdots 0)}^{\pupsq{\textrm{SU}(n)_k}}
	-
	h_{(0 1 0 \cdots 0)}^{\pupsq{\textrm{SU}(k)_n}}
	=
	\frac{2}{n k}.
\end{align}
Restoring $\lambda_A > 0$, the conformal dimension of the interaction in the U$(1) \, \oplus \, $ SU$(n)_k$ 
remnant theory is
\begin{align}\label{hBDef}
	h^{(\msf{B})} 	
	=
	\frac{2}{n k(1 + \lambda_A n k)}
	+
	\frac{(n + 2)(n-1)}{n(n + k)}.
\end{align}
We can write the corresponding diagonal interaction operator as
\begin{align}\label{OB}
	\mathcal{O}_{a}^{(\msf{B})}
	\equiv&\,
	I_{[v_1 v_2] a} \bar{I}^{\dagger \, [v_2 v_1] a},
	\quad
	x_2^{(\msf{B})}
	=
	\frac{2(k+2)}{k^2}
	- 
	4 \lambda_A,
\end{align}
where we have taken the replica limit of twice Eq.~(\ref{hBDef}), and 
where we have defined
\begin{align}
	I_{[v_1 v_2] a} \equiv L_{v_1,a} L_{v_2,a}, \;\;
	\bar{I}^{\dagger \, [v_1 v_2] a} \equiv R^{\dagger v_1, a} R^{\dagger v_2, a}.
\end{align}

Interaction $\mathcal{O}_{a}^{(\msf{C})}$ carries the same scaling dimension,
and is defined via
\begin{align}\label{OC}
	\mathcal{O}_{a}^{(\msf{C})}
	\equiv&\,
	I^{\dagger \, [v_1 v_2] a} \bar{I}_{[v_2 v_1] a},
	\quad
	x_2^{(\msf{C})}
	=
	\frac{2(k+2)}{k^2}
	- 
	4 \lambda_A,	
\end{align}
where
\begin{align}
	I^{\dagger \, [v_1 v_2] a} \equiv L^{\dagger\, v_1, a} L^{\dagger \, v_2, a},\;\;
	\bar{I}_{[v_1 v_2] a} \equiv R_{v_1, a} R_{v_2, a}.
\end{align}

{\flushleft{\underline{AIII interactions and Fierz identities}}}\\

We write the interaction Hamiltonian as a generic Hermitian linear combination of the operators
in Eqs.~(\ref{OA1}), (\ref{OA2}), (\ref{OB}), and (\ref{OC}):
\begin{align}\label{haiiNoFierz}
	\haii = \sum_{a}
	\int
	d^2\vex{r} 
	&
	\left[
	U \mathcal{O}_{a}^{(\msf{A},1)} + V \mathcal{O}_{a}^{(\msf{A},2)} 
	\right.
	\nonumber\\
	&\,
	\left.
	+ W \left(\mathcal{O}_{a}^{(\msf{B})} + \mathcal{O}_{a}^{(\msf{C})}\right) 
	\right].
\end{align}
We exploit the SU$(k)$ Fierz identity in Table~\ref{FierzTable} to
rewrite
\begin{align}\label{OA1Def}
	\mathcal{O}_{a}^{(\msf{A},1)}
	=
	-
	2 (L^{\dagger}_{a} R_a) (R^\dagger_{a} L_a)  
	-
	\frac{2}{k} J_a \bar{J}_a.
\end{align}
We also have
\begin{align}\label{OBCDef}
	\mathcal{O}_{a}^{(\msf{B})}
	=
	(R^\dagger_{a} L_a)^2,
	\;\;
	\mathcal{O}_{a}^{(\msf{C})}
	=
	(L^\dagger_{a} R_a)^2.
\end{align}
Clearly, for the case of a single species (``valley'') $k = 1$,
$\mathcal{O}_{a}^{(\msf{A},1)} = \mathcal{O}_{a}^{(\msf{B})} = \mathcal{O}_{a}^{(\msf{C})} = 0$.

The spin density $S^z$ and the spin singlet, valley singlet mass operator $m$
were defined in Eqs.~(\ref{AbelCurr}) and (\ref{mass}).
Expressing these via the holomorphic decomposition in Eq.~(\ref{AIIIDecomp}) gives
\bsub\label{SzmDefs}
\begin{align}
	S^z_a 
	=&\, 
	\frac{1}{2} \left(R_a^\dagger L_a + L_a^\dagger R_a\right),
	\\
	m_a 
	=&\,
	R^\dagger_a L_a - L^\dagger_a R_a,
\end{align}
\begin{align}
	\Rightarrow
	R_a^\dagger L_a
	=
	S^z_a + \frac{1}{2} m_a, 
	\;\;
	L^\dagger_{a} R_a
	=&\,
	S^z_a - \frac{1}{2} m_a. 
\end{align}
\esub
We then arrive at
\begin{align}\label{AIIIFierzRes}
\begin{aligned}
	\mathcal{O}_{a}^{(\msf{A},1)}
	=&\,
	\frac{1}{2}
	\left(
	m_a m_a
	-
	4
	S^z_a S^z_a
	\right)
	-
	\frac{2}{k}
	J_a \bar{J}_a,
	\\
	\mathcal{O}_{a}^{(\msf{B})} + \mathcal{O}_{a}^{(\msf{C})}
	=&\,
	\frac{1}{2}
	\left(
	m_a m_a 
	+
	4 S^z_a S^z_a  
	\right).
\end{aligned}
\end{align}
Using Eqs.~(\ref{OA2}) and (\ref{AIIIFierzRes}) in Eq.~(\ref{haiiNoFierz}),
we recover the form of the class AIII interaction Hamiltonian
in Eq.~(\ref{haii}).

The ``tree level'' class AIII RG flows in Eq.~(\ref{AIIITreeFlow}) obtain from 
inserting $x_1$ from Eq.~(\ref{LDoSAIII-CFT}) and the interaction 
dimensions from Eqs.~(\ref{OA1}), (\ref{OA2}), (\ref{OB}) and (\ref{OC}) 
into Eq.~(\ref{UFlowFrame}).


\subsection{Class DIII}

\subsubsection{Density of states and multifractal spectrum \label{Sec: CFTDIII-MFC}}

In class DIII, the effective CFT is SO$(n)_k$ [Eq.~(\ref{Embed})]. 
In fact, it is convenient to consider $2 n$ replicas, so that we work
with the affine version of the $D_n$ Lie algebra. 
As the results for LDoS scaling dimensions are new to our knowledge,
we provide more details than we did in Sec.~\ref{Sec: CFTAIII-MFC}, above. 

We will denote the (regular, non-affine) fundamental weights
of the algebra as $\omega_p$, $p \in \{1,2,\ldots,n\}$. 
Affine weights will be denoted with a ``hat,''
$\hat{\omega}_m$, $m \in \{0,1,2,\ldots,n\}$.
The following facts about the quadratic form matrix\cite{CFT} 
$F_{i,j} \equiv \langle\omega_i,\omega_j\rangle$ are useful:
\begin{align}
\begin{aligned}
	F_{i,j} 
	=&\, 
	\min(i,j), \;\; 1 \leq \{i,j\} \leq n - 2,
	\\
	F_{i,n-1} 
	=&\,
	F_{i,n}
	=
	\frac{i}{2}, \;\; 1 \leq i \leq n - 2,
	\\
	F_{n-1,n}
	=&\,
	\frac{n-2}{4},\quad
	F_{n-1,n-1} 
	=
	F_{n,n}
	=
	\frac{n}{4}.
\end{aligned}
\end{align}
The ``outer'' [$(n-1)^{\text{th}}$ and $n^{\text{th}}$] rows and columns reflect the
two spinor representations. 
These results are most easily obtained using the expansion of the fundamental 
weights in terms of the orthonormal weights of the fundamental representation.

We consider first the minimal non-abelian case with $k = 3$ valleys.
At level one (free Majorana fermions), only the $\omega_1$, $\omega_{n-1}$ and $\omega_n$
representations correspond to primary fields. At level $k = 2$, the full set consists of the affine weights
\begin{align}\label{level 2 weights DIII}
\begin{aligned}
	2 \hat{\omega}_0 \rightarrow&\, [2 \, 0 \, 0 \, 0 \cdots 0 \, 0] \Leftrightarrow \mathbb{I},
	\\
	\\
	\hat{\omega}_0 + \hat{\omega}_1 \rightarrow&\, [1 \, 1 \, 0 \, 0 \cdots 0 \, 0] \Leftrightarrow \Omega_i,
	\\
	\hat{\omega}_0 + \hat{\omega}_{n-1} \rightarrow&\, [1 \, 0 \, 0 \, 0 \cdots 1 \, 0] \Leftrightarrow \Omega_{\sigma},
	\\
	\hat{\omega}_0 + \hat{\omega}_{n} \rightarrow&\, [1 \, 0 \, 0 \, 0 \cdots 0 \, 1] \Leftrightarrow \Omega_{\bar{\sigma}},
	\\
	\\
	2 \hat{\omega}_1 \rightarrow&\, [0 \, 2 \, 0 \, 0 \cdots 0 \, 0] \Leftrightarrow \tilde{\Omega}_{(i j)},
	\\
	2 \hat{\omega}_{n-1} \rightarrow&\, [0 \, 0 \, 0  \cdots  0 \, 2 \, 0],\quad 2 \hat{\omega}_{n} \rightarrow [0 \, 0 \, 0  \cdots  0 \, 0 \, 2]  
	\\	&\, \quad\Leftrightarrow \Omega_{[i_1 i_2 \cdots i_{n}]},
	\\
	\\
	\hat{\omega}_1 + \hat{\omega}_{n-1} \rightarrow&\, [0 \, 1 \, 0 \, 0 \cdots 1 \, 0] \Leftrightarrow \Omega_{i \sigma}
	\\
	\hat{\omega}_1 + \hat{\omega}_{n} \rightarrow&\, [0 \, 1 \, 0 \, 0 \cdots 0 \, 1] \Leftrightarrow \Omega_{i \bar{\sigma}}
	\\
	\hat{\omega}_{n-1} + \hat{\omega}_{n} \rightarrow&\, [0 \, 0 \, 0 \, 0 \cdots 1 \, 1] \Leftrightarrow \Omega_{[i_1 i_2 \cdots i_{n-1}]},
	\\
	\\
	\hat{\omega}_2 \rightarrow&\, [0 \, 0 \, 1 \, 0 \cdots 0 \, 0] \Leftrightarrow \Omega_{[i j]},
	\\
	\hat{\omega}_3 \rightarrow&\, [0 \, 0 \, 0 \, 1 \cdots 0 \, 0] \Leftrightarrow \Omega_{[i j k]},
	\\
	\vdots
	\\
	\hat{\omega}_{n-2} \rightarrow&\, [0 \, 0 \, 0 \cdots 1 \, 0 \, 0] \Leftrightarrow \Omega_{[i_1 i_2 \cdots i_{n-2}]}.
\end{aligned}	
\end{align}
Here the numbers in square brackets are affine Dynkin coefficients. The affine weight $\hat{\omega}_0$ is the
basic fundamental weight (i.e., the vacuum for $k = 1$). 
The $\Omega$'s denote the associated holomorphic primary fields, irreducible tensors with indices that transform in some
combination of fundamental replica ($i,j \in \{1,2,\ldots,2n\}$) and spinor ($\sigma,\bar{\sigma}\in \{1,2,\ldots,2^{n-1}\}$) representations. 
In particular, $\Omega_{[i_1 i_2 \cdots i_{p}]}$ is a fully antisymmetric rank-$p$ tensor,
while $\tilde{\Omega}_{(i j)}$ is the traceless symmetric 2nd rank tensor. 

All of the fields in Eq.~(\ref{level 2 weights DIII}) appear as primaries at level $k = 3$, since we can add $\hat{\omega}_0$ 
to each. Additional dominant weights at level 3 are the unique triple sums of 
$\{\hat{\omega}_1,\hat{\omega}_{n-1},\hat{\omega}_{n}\}$, and
the sum of one of these with one of the level 2 weights $\{\hat{\omega}_2,\hat{\omega}_3,\ldots,\hat{\omega}_{n-2}\}$.
In particular, we get mixed tensors such as
\begin{gather}
	\hat{\omega}_1 + \hat{\omega}_i \rightarrow [0 \, 1 \, 0 \, 0 \cdots 0 \, \stackrel{i^{\mathrm{th}} \mathrm{ place}}1 \, 0 \cdots 0 \, 0]
	\Rightarrow 
	\Omega_{j' [j_1 j_2 \cdots j_{i}]}, 
	\nonumber\\
	\quad i \in \{2,3,\ldots,n-2\}.
\end{gather}
The weights $\hat{\omega}_{n-1} + \hat{\omega}_i$ and $\hat{\omega}_{n} + \hat{\omega}_i$ belong
to the conjugacy classes of the spinor representations.

The $q{\mathrm{th}}$ LDoS moment involves exactly $q$ distinct replica labels (no more, no less). 
Therefore, this should be associated to the most relevant component of the decomposition 
of a generic tensor $T_{i_1 i_2 \cdots i_q}$ ($q \ll 2 n$). 
This rules out the basic spinor (twist field) representations $\hat{\omega}_{n-1}$ and $\hat{\omega}_n$,
and all fields in their conjugacy classes. 
At level $k = 3$, for the $q$th moment there are only two possibilities:
\begin{align}
	\hat{\omega}_0 + \hat{\omega}_q:
	\quad
	h_{\omega_q}
	=&\, 
	\frac{\langle\omega_q,\omega_q + 2 \rho\rangle}{2(k+g)}
	=
	\frac{F_{q,q} + 2 \sum_{i = 1}^{n} F_{q,i}}{2(k + 2n - 2)}	
	\nonumber\\
	=&\,
	\frac{(2 n - q)q}{2(k + 2n - 2)}.	
	\label{T[]DIII}
\end{align}
Note that we recover the free fermion dimension $h = 1/2$ for $k = q = 1$ (and any $n$).
\begin{align}
	\hat{\omega}_1 + \hat{\omega}_{q - 1}:
	\quad
	h_{\omega_1 + \omega_{q-1}}
	=&\,
	\frac{
	\left[
	\begin{aligned}
	&\,
	F_{q-1,q-1} 
	+ 
	2 F_{q-1,1} 
	+ 
	F_{1,1}
	\\&\,
	+ 
	2 \sum_{i = 1}^{n} (F_{1,i} + F_{q-1,i})
	\end{aligned}
	\right]
	}{2(k + 2n - 2)}	
	\nonumber\\
	=&\,
	\frac{(2 + 2 n - q) q}{2(k + 2n - 2)}.	
\end{align}
We see that the antisymmetric tensor $\Omega_{[i_1 \cdots i_q]} \Leftrightarrow \omega_q$ 
corresponds to the most negative scaling dimension in the replica $n \rightarrow 0$ limit. 
The same applies in classes AIII and CI, and holds for larger $k$ as well.   

Thus, the scaling dimension of the $q$th LDoS moment is given by 
\begin{align}\label{DeltaqDIII-CFT}
	x_q
	=&\,
	2 h_{\omega_q}
	=
	-
	\frac{q^2}{(k - 2)},
\end{align}
where we have taken the replica limit. 
The class DIII multifractal spectrum in Eq.~(\ref{MFCSpec}) and (\ref{MFCDim}) obtains from 
combining Eqs.~(\ref{tauqDefLDoS}) and (\ref{DeltaqDIII-CFT}).

We note that the $k = 3$ case corresponds to quite strong multifractality: 
the critical $q$ for multifractal spectral termination\cite{Chamon96,Castillo97,Carpentier01,Foster09} 
is given by
\begin{align}
	q_c = \sqrt{2 (k - 2)} = \sqrt{2}. 
\end{align}
Thus, unlike the minimal non-abelian realizations of classes CI and AIII, the second moment 
is already beyond termination. For $k = 3$, this potentially complicates the treatment of interactions (four-fermion operators),
because annealed (disorder) and quenched (spatial) averages are no longer the same.\cite{Chamon96,Foster09}
However, it turns out that interaction operator scaling dimension $x_2^{(U)} = 0 > x_2$
(Secs.~\ref{DIIIResults} and \ref{Sec: CFTDIII-Int}), so that this issue does not arise.

\subsubsection{Interaction operator \label{Sec: CFTDIII-Int}}

The mass-squared (BCS pairing) interaction operator in Eq.~(\ref{hdii}) can be
expressed through the decomposition in Eq.~(\ref{DIIIDecomp}) as
\begin{align}
	m_a m_a
	=&\,
	-4 L_{v_1,a} R_{v_1,a} L_{v_2,a} R_{v_2,a}
	=
	-2 J_{\kappa a}^i \bar{J}_{\kappa a}^i, 
\end{align}
where the replica-resolved valley SO$(k)$ currents are defined via
\begin{align}
	J_{\kappa a}^{i} \equiv L_{a} \tk^i L_{a},
	\;\;
	\bar{J}_{\kappa a}^{i} \equiv R_{a} \tk^i R_{a},
\end{align}
and we have used the Fierz identity in Table~\ref{FierzTable}.	
The holomorphic field $J_{\kappa a}^{i}$ transforms in the 
antisymmetric adjoint representation ($\omega_2$) in valley space, and
the traceless symmetric representation ($2 \omega_1$) in replica space. 
Here, $\omega_p$ denotes the $p$th (non-affine) fundamental weight. 
In the remnant SO$(2n)_k$ theory, this has the holomorphic dimension
\begin{align}\label{DIIIIntHoloDim}
	h_{(2 0 0 \cdots 0)} = 
	\frac{4 F_{1,1} + 4 \sum_{i = 1}^{n} F_{1,i}}{2(k + 2n - 2)}	
	=&\,
	\frac{2 n}{(k + 2n - 2)}.	
\end{align}
As a check, for $k$ even the dimension of the antisymmetric representation in the 
SO$(k)_{2n}$ theory is [Eq.~(\ref{T[]DIII})]
\begin{align}
	h_{(0 1 0 \cdots 0)} = 
	\frac{2(k - 2)}{2(k + 2n - 2)},
\end{align}
so that $h_{(2 0 0 \cdots 0)}+h_{(0 1 0 \cdots 0)} = 1$.
Taking the replica limit of Eq.~(\ref{DIIIIntHoloDim}) gives $x_2^{(U)} = 0$ for
the interaction in Eq.~(\ref{hdii}). The flow equation (\ref{DIIITreeFlowU})
then obtains from Eq.~(\ref{UFlowFrame}) and (\ref{DeltaqDIII-CFT}) with $q = 1$.


\section{Conclusions and open questions \label{Sec: Conc}}

In summary, surface states of topological superconductors (TSCs) are critically
delocalized in the absence of interactions, with wavefunction statistics determined
exactly by conformal field theory (CFT).\cite{Ludwig94,Nersesyan94,Mudry96,Caux96,Chamon96,Ludwig00,Bhaseen01,LeClair08,Foster12-B} 
Clean surface states are robust against sufficiently weak interactions, owing to the vanishing density of 
states at the Majorana surface band Dirac point. 

The main result of this paper is that the combination of \emph{both} weak disorder \emph{and} 
weak interactions is more subtle than either in isolation. Wavefunction multifractality\cite{MFCRev,Loc} 
(spatial inhomogeneity) and Chalker scaling\cite{Chalker88,Chalker90,Cuevas07} (correlations between the spatial structures of 
different wavefunctions) can amplify interparticle interactions.\cite{Feigelman07,Feigelman10,Foster12-B} 
For class CI, the effect is so strong that we predict the absence of gapless surface states, as the multifractal enhancement 
of interaction matrix elements destabilizes the non-interacting fixed point.\cite{Foster12-B} 
The Wess-Zumino-Novikov-Witten\cite{Witten84} Finkel'stein non-linear sigma model\cite{Finkelstein,BK} 
(WZNW-FNLsM) does not reveal the existence of a perturbatively accessible critical fixed point, 
and the most likely scenario is that time-reversal symmetry always breaks spontaneously. The dominant 
instability realizes the (class C) spin quantum Hall effect at the surface.\cite{SpinQHE,SRL09,Foster12-B}

By contrast, stable surface states in the presence of both disorder and interactions are
possible in classes AIII and DIII. For a class AIII topological superconductor with a 
winding number modulus $|\nu| \geq 2$, CFT predicts a window of disorder strengths in which 
interactions are irrelevant. By combining the CFT scaling dimensions with a phenomenological BCS term 
that favors pairing of the \emph{surface} quasiparticles,
we predict interaction-stabilized critical fixed points outside of this range. 
For $|\nu| \gg 1$, this agrees with the one-loop result of the WZNW-FNLsM, which is well-controlled
in this limit. The stability of these non-trivial fixed points to higher order loop 
corrections is an important topic for future work, as is the characterization 
of observables. 

In class DIII, CFT predicts that interactions are always irrelevant. This 
result obtains despite the fact that the disorder-averaged density of states diverges near 
zero energy in class DIII, for $|\nu| \geq 3$. The key is that there is no multifractal 
enhancement of the interaction matrix elements in this case, due to the absence of a continuous symmetry
in the presence of disorder. The surface states of both classes AIII and DIII can be
gapped by sufficiently strong interactions, leading to spontaneous time-reversal
symmetry breaking. The instabilities are expected to produce surface spin or thermal 
quantum Hall states.\cite{Ryu12,Stone12} 

In the absence of interactions, the dc zero temperature spin conductance is universal 
at the surface of topological superconductors in classes CI and AIII.\cite{Ludwig94,Tsvelik95,Ostrovsky06} 
We argued that the 
surface thermal conductance in class DIII is also universal.
We discussed (but did not derive) the result that the lowest-order Altshuler-Aronov 
interaction conductance corrections vanish in the conformal limit. 
We derive and examine this point in detail elsewhere.\cite{WZNW-P3}
The full one-loop WZNW-FNLsM RG equations are also derived in that work.

In Ref.~\onlinecite{YZC-P1}, we verified CFT predictions for non-interacting
surface states with two valleys in classes CI and AIII using numerical methods. 
In that work, we considered the average density of states and the multifractal spectrum of local density of states
fluctuations.
To our knowledge, transport has not been simulated in these models. 
Class DIII for $|\nu| \geq 3$ remains open for numerical investigation.
In particular, we argued here that the thermal conductance is universal, and
that this is reflected by the topological protection of conformal invariance.
CFT gives universal predictions for multifractal spectra and the dynamic critical exponent. 
On the other hand, it is known that one can realize a diffusive ``thermal metal''
phase in class DIII.\cite{Senthil00} While our arguments suggest that this does not occur
at the surface of a topological superconductor, it would be interesting to 
verify this result with numerics. 
Another non-trivial check would be to study surface states in disordered 3D
lattice models of bulk TSCs.


\begin{acknowledgments}

We thank V.\ Kravtsov, M.\ Mueller, A.\ Scardicchio, I. Gornyi, I. Gruzberg, and A. Mirlin
for helpful discussions.
This work was supported by the Welch Foundation under Grant No.~C-1809.

\end{acknowledgments}

\appendix

\section{Non-valley invariant mass operators: class CI \label{App: NonInvtMass}}

In addition to the unique valley- and spin-symmetric mass
operator defined in Eq.~(\ref{mass}), there are other Dirac mass
operators that do not preserve valley and/or spin symmetry. 
These also break time-reversal invariance. 
The non-invariant mass operators can be organized into different 
irreducible representations of the valley and spin symmetry groups. 
In class CI, for example, there are two such classes of operators:
\bsub
\begin{align}
	M^a(\vex{r}) \equiv&\, \psi^\dagger \sigh^3 \hat{\mathfrak{a}}_\mu^a \kah^\mu \psi,
	\label{CImassV}
	\\
	\mathscr{M}^{n,i}(\vex{r}) \equiv&\, \LF \, \hat{s}^n \tk^i \, \rt.
	\label{CImassVS}
\end{align}
\esub
In Eq.~(\ref{CImassV}),  $\mu \in \{0,1,2,3\}$ and $\kah^\mu = \{\mathbb{I},\kah^{1,2,3}\}$ 
are the $2 k \times 2 k$ block identity and Pauli matrices in valley space.
The latter commute with the $k \times k$ matrices $\hat{\mathfrak{a}}_\mu^a$ which 
satisfy 
\[
	\tr[\hat{\mathfrak{a}}_0^a] = 0, 
	\;\;
	\left(\hat{\mathfrak{a}}_0^a\right)^\T = \hat{\mathfrak{a}}_0^a,
	\;\;
	\left(\hat{\mathfrak{a}}_{1,2,3}^a\right)^\T = -\hat{\mathfrak{a}}_{1,2,3}^a.
\]	
We have expressed $\mathscr{M}^{n,i}$ in terms of the decomposition in 
Eq.~(\ref{CIDecomp}); $\hat{s}^n$ denotes a spin space Pauli matrix. 

The mass operator $M^a$ is spin SU$(2)$ invariant, and transforms 
under the $\omega_2$ (antisymmetric second rank tensor) representation 
of the valley Sp$(2 k)$ symmetry group. (Here $\omega_2$ is the second
fundamental weight.\cite{CFT}) A non-zero $\langle M^a \rangle$ allows
the realization of spin Hall plateaux with $k$ replaced by any 
$p \in \{-k, -k + 2, \ldots, k - 2, k\}$ in Eq.~(\ref{SpinQHECond}). 

The mass operator $\mathscr{M}^{n,i}$ transforms in the  
adjoint representations of the spin SU(2) and valley Sp$(2 k)$ symmetries. 
A non-zero average of this operator would imply broken spin symmetry,
and the resulting surface quantum Hall state will reside in either
class A [residual spin U(1) invariance] or class D (no spin symmetry).

\section{Flow equation for $\lambda_A$ in class AIII away from the conformal fixed point \label{App: lambdaAFlow}}

In the absence of interparticle interactions, the action for the U$(1) \oplus \text{SU}(n)_k$ class AIII 
low-energy effective field theory is transcribed in Eq.~(\ref{AIIIWZNW}), above. 
For a non-topological (Gade) class AIII disordered quantum
system in 2D, the replica theory for quantum diffusion
(principal chiral model) is  
\begin{align}\label{SPCM}
\begin{aligned}
	S_{\mathrm{AIII}}^{\puprm{PCM}}
	=&\,
	\frac{1}{\lambda}
	\int
	\frac{d^2\vex{r}}{8 \pi}
	\tr\left[
	\Nabla \hat{U}^\dagger
	\cdot
	\Nabla \hat{U}
	\right]
	\\&\,
	-
	\frac{\lambda_A}{\lambda^2}
	\int
	\frac{d^2\vex{r}}{8 \pi}
	\left[
	\tr
	\left(
	\hat{U}^\dagger 
	\Nabla \hat{U}
	\right)
	\right]^2
\end{aligned}
\end{align}
where $\hat{U}$ is a U$(n)$-valued matrix field.
We write
\[
	\hat{U} \equiv \exp(i \phi) \, \hat{Q},
\]
and thereby obtain
\begin{align}
\begin{aligned}
	S_{\mathrm{AIII}}^{\puprm{PCM}}
	=&\,
	\frac{n}{\lambda}
	\left(
	1
	+
	\frac{n \lambda_A}{\lambda}
	\right)
	\int
	\frac{d^2\vex{r}}{8 \pi}
	\Nabla \phi \cdot \Nabla \phi	
	\\&\,
	+
	\frac{1}{\lambda}
	\int
	\frac{d^2\vex{r}}{8 \pi}
	\tr\left[
	\Nabla \hat{Q}^{\dagger}
	\cdot
	\Nabla \hat{Q}
	\right].
\end{aligned}
\end{align}
The coupling strength of the $U(1)$ term (free boson) cannot be renormalized,
so that 
\begin{align}\label{NoU1Renorm}
	\frac{d \lambda_A}{d l}
	=
	\left(
	\frac{1}{n}
	+
	2
	\frac{\lambda_A}{\lambda}
	\right)
	\frac{d \lambda}{d l}.
\end{align}

Eq.~(\ref{SPCM}) is a consistent deformation of the AIII model away from
$\lambda = 1/k$ [Eq.~(\ref{SWZNW})], except that it is missing the 
Wess-Zumino-Novikov-Witten term. 
Incorporating the latter, the one-loop beta equation for $\lambda$ 
is given by 
\begin{align}\label{lambdaFlowAIIIn}
	\frac{d \lambda}{d l} = n \lambda^2 \left[1 - (k \lambda)^2\right] + \ldots,
\end{align}
where we have not yet taken the replica $n \rightarrow 0$ limit.
Combining Eqs.~(\ref{NoU1Renorm}) and (\ref{lambdaFlowAIIIn})
leads to
\begin{align}
	\label{lambdaA WZNWFlow}
	\frac{d \lambda_A}{d l}
	=&\,
	\lambda^2 \left[1 - (k \lambda)^2\right],
\end{align}
Eq.~(\ref{lambdaA WZNWFlow}) was first obtained in Ref.~\onlinecite{GLL},
wherein the authors claimed that this is the \emph{exact} beta function 
for the perturbation $\lambda \neq 1/k$ of the class AIII CFT, in the absence of interactions.
It implies that 
\begin{itemize}
\item{$\lambda_A$ is always generated under the RG \emph{except} at the WZNW point $\lambda = 1/k$, and}
\item{$\lambda_A$ is driven to arbitrarily positive (negative) values for $\lambda < 1/k$ ($\lambda > 1/k$).
} 
\end{itemize}

\section{Tree level RG for interactions at a critically delocalized fixed point \label{App: TreeFlow}}

In this Appendix we derive Eq.~(\ref{UFlowFrame}).
Consider the class CI theory in Eq.~(\ref{hci}).
In a fixed realization of the disorder, we can write a (2+1)-D zero-temperature 
imaginary time path integral with the action
\begin{equation}
	\label{App--SCI}
	\begin{gathered}
	S
	=
	\int
	d \tau \,
	d^2 \vex{r} \,
	\bar{\psi}
	\left\{
	\eta
	\partial_\tau
	+
	\frac{1}{2 \pi}
	\left[
	\sigb \cdot
	\left(
	i \Nabla
	-
	\vex{A}_i \, 
	\tk^i
	\right)
	\right]
	\right\}
	\psi
	\\
	\;\;
	+
	U
	\sum_{a}
	T^{{\scriptscriptstyle{\{\sigma_i\}}}}_{{\scriptscriptstyle{\{v_i\}}}}
	\!\!
	\int
	d \tau \,
	d^2 \vex{r} \,
	\bar{\psi}_{\sigma_1,v_1,a}
	\bar{\psi}_{\sigma_2,v_2,a}
	\psi_{\sigma_3,v_3,a}
	\psi_{\sigma_4,v_4,a},
\end{gathered}
\end{equation}
where we have included a generic four-fermion interaction
with coupling strength $U$.
Here $a$ denotes the replica index, while the pseudospin-valley tensor
$T^{{\scriptscriptstyle{\{\sigma_i\}}}}_{{\scriptscriptstyle{\{v_i\}}}}$ 
is chosen such that the interaction respects
the defining class CI symmetries [time reversal and spin SU(2) invariances]. 
The parameter $\eta$ multiplying the imaginary time derivative
is necessary to keep track of the dynamic critical exponent, i.e.\
the fact that time and space need not scale in the same way 
in the disorder-averaged theory. 
The bare value of $\eta = 1$, but under renormalization
this parameter will flow. We use a renormalization group (RG) 
scheme that fixes the coefficient of the kinetic term
$\bar{\psi} i \sigb \cdot \Nabla \psi$. 
An alternative would be to fix the $\bar{\psi} \partial_\tau \psi$ 
term and renormalize the Fermi velocity.\cite{Foster08-Graphene}
This is however undesirable here, as it will lead to a renormalization
of the spatial components of the stress tensor. We require a fixed normalization 
of the latter to determine scaling dimensions using (2+0)-D conformal field theory.\cite{CFT}

We assume that the imaginary time variable $\tau$ carries units of $L^z$, where 
$L$ denotes the system size and $z$ is a (possibly scale-dependent)
``dynamic critical exponent.'' In inverse length units, one then has the dimensions
\begin{align}
	\left[\psi(\tau,\vex{r}) \right] = \frac{1+z}{2},\;\;	
	\left[\eta\right] = 1 - z,\;\;
	\left[U\right] = - z.
\end{align}
The tree level RG equation for $U$ is then given by 
\begin{align}\label{App--UFlowFrame1}	
	\frac{d \ln U}{d l} 
	=
	-z
	+
	\left(\frac{d \ln U}{d l}\right)_{a}, 
\end{align}	
where the second term is the ``anomalous'' dimension due to renormalization by the disorder. 

Averaging over disorder as in Eq.~(\ref{VecVarNA}) induces an effective four-fermion disorder
vertex. The renormalization of $U$ due to the disorder involves an infinite sum over diagrams
with one interaction vertex and any number of disorder vertices. Because the disorder is static,
these diagrams do not involve integrals over Matsubara frequencies; the ultraviolet divergences
obtain exclusively from momentum integrations.
Given the strong correlations in spatial structure implied by Chalker scaling\cite{Chalker88,Chalker90,Cuevas07} 
for the low energy wavefunctions, it is sufficient to determine the renormalization
of the interaction matrix elements in an effective (2+0)-D theory. This can be understood
as the projection of Eq.~(\ref{App--SCI}) to zero Matsubara frequency for all fermion fields.
In two spatial dimensions, a four-fermion term has engineering dimension equal to two; therefore 
\begin{align}\label{App--UFlowFrame2}
	\left(\frac{d \ln U}{d l}\right)_{a} = 2 - x_2^{(U)},
\end{align}
where $x_2^{(U)}$ is the scaling dimension of the four-fermion interaction operator in the
disorder-averaged (2+0)-D theory.

The dynamic critical exponent $z$ for the disorder-averaged non-interacting model at zero
temperature is uniquely determined by the critical behavior of the density of states (DoS). 
In particular, the DoS scaling dimension $x_1 = 2 - z$ [Eq.~(\ref{zDoS})].
Combining this with Eqs.~(\ref{App--UFlowFrame1}) and (\ref{App--UFlowFrame2}) gives Eq.~(\ref{UFlowFrame}).

\section{Class AIII WZNW-FNLsM interaction-stabilized non-trivial fixed point \label{App: AIIIFP}}

The interaction flow equations (\ref{tFlowAIII}) and (\ref{cFlowAIII}) possess a non-trivial
fixed point. Setting $\lambda = 1/k$ and working to order 
$\{\gamma_c/k,\gamma_c \lambda_A, \gamma_c^2\}$, the fixed point is located at 
\begin{align}\label{AIIIFPFull}
\begin{aligned}
	\gamma_t^*
	=&\,
	\frac{\lambda_A (2 k \lambda_A-1) (3 k \lambda_A-2)}{2 (k \lambda_A-1) \left[\lambda_A (3 k \lambda_A+k-2)-1\right]},
	\\
	\gamma_c^*
	=&\,
	\frac{\lambda_A (2-3 k \lambda_A)}{2(k \lambda_A - 1)}.
\end{aligned}
\end{align}


\begin{thebibliography}{99}
\bibitem{footnote--BDI}
	In Fig.~\ref{Fig--ChalkerWF} we plot 2D wavefunctions obtained by exact diagonalization
	of a class BDI nearest-neighbor random hopping model on the square lattice. 
	The size of the system is $44 \times 44$, with periodic boundary conditions. 
	This is a non-topological system that exhibits critically delocalized states
	near zero energy, due to an artificial chiral symmetry associated to the special form of the disorder.\cite{Loc} 
	The probability densities for the first four positive energy states are shown in a), 
	while 4 successive localized states near the upper band edge are shown in b).
	The surface states of a class CI topological superconductor have the same 
	qualitative features: critical delocalization at zero energy, and localization
	away from this. (The crucial difference is that the special form of the disorder 
	in the latter case is determined by the topology, rather than by fine-tuning on the
	lattice scale.) We plot wavefunctions for the 2D BDI model because they entail
	less computation than class CI surface states. To obtain the latter, one must
	in principle diagonalize a surface-disordered 3D bulk model\cite{SRL09,Schubert12} 
	(but see e.g.\ Ref.~\onlinecite{YZC-P1}).
\bibitem{footnote--LDoSOP}
	In a (2+1)-D formulation of the theory, the density of states operator in imaginary 
	time is local in space and Matsubara frequencies:
	$\nu(\e,\vex{r}) \sim \psi^\dagger(\e_n,\vex{r}) \psi(\e_n,\vex{r})$.
\bibitem{footnote--PCMWZNW}
	The theory with $\lambda \neq 1/k$ should be referred to as a principal 
	chiral model plus a Wess-Zumino-Novikov-Witten term.
\bibitem{footnote--gammac}
	In principle, one can obtain results to higher orders in 
	the Cooper channel interaction $\gamma_c$ in a one-loop 
	Finkel'stein sigma model calculation. However, substantial 
	technical difficulties arise,\cite{BK} and for this reason one 
	typically works only to the lowest non-trivial order.
\end{thebibliography}
\end{document}